\documentstyle[12pt]{article}
\let\ssection=\section
\renewcommand{\section}{\setcounter{equation}{0}\ssection}

\newcommand{\be}{\begin{enumerate}}
\newcommand{\ee}{\end{enumerate}}
\newcommand{\bi}{\begin{itemize}}
\newcommand{\ei}{\end{itemize}}
\newcommand{\bd}{\begin{description}}
\newcommand{\ed}{\end{description}}
\newcommand{\beq}{\begin{equation}}
\newcommand{\eeq}{\end{equation}}
\newcommand{\beqn}{\begin{eqnarray}}
\newcommand{\eeqn}{\end{eqnarray}}

\newcommand{\cf}{{\em cf\ }}

\newcommand{\ie}{{\em i.e.,\ }}
\newcommand{\etc}{{\em etc }}
\newcommand{\Kuchar}{Kucha\v{r}\ }

\newcommand{\Alg}{Alg}

\newcommand{\ad}{{\rm ad}}
\renewcommand{\bar}[1]{\overline{#1}}           
\newcommand{\bra}[1]{\langle#1|}
\renewcommand{\c}{\circ}                        
\newcommand{\half}{{\frac 12}}
\newcommand{\k}{\vec k}
\newcommand{\ket}[1]{|#1\rangle}
\newcommand{\la}{\langle}
\newcommand{\ra}{\rangle}
\newcommand{\map}{\longrightarrow}
\newcommand{\norm}[1]{\|\,{#1}\,\|}
\newcommand{\op}[1]{\widehat{#1}}
\newcommand{\twid}[1]{\tilde{#1}}
\newcommand{\unit}{1}

\newcommand{\x}{\vec x}

\renewcommand{\[}{[\,}                          
\renewcommand{\]}{\,]}                          

\newcommand{\A}{{\cal A}}
\newcommand{\Ad}{{\rm Ad}}
\newcommand{\C}{\mkern1mu\raise2.2pt\hbox{$\scriptscriptstyle|$}
  {\mkern-7mu\rm C}}                            
\newcommand{\D}{\Delta}
\newcommand{\G}{{\cal G}}
\newcommand{\Ga}{\Gamma}
\renewcommand{\H}{{\cal H}}                     
\newcommand{\HA}{{\cal H}{\cal A}}
\renewcommand{\L}{{\cal L}}                     
\newcommand{\LS}{{\cal L}_\Sigma}
\newcommand{\Lx}{{\cal L}_{x_0}}
\newcommand{\Om}{\Omega}
\newcommand{\R}{{\rm I\! R}}                    
\newcommand{\Si}{\Sigma}
\newcommand{\Th}{\Theta}
\newcommand{\Tr}{{\rm Tr\,}}

\renewcommand{\a}{\alpha}                       
\renewcommand{\b}{\beta}                        
\newcommand{\g}{\gamma}
\newcommand{\de}{\delta}
\newcommand{\e}{\epsilon}
\renewcommand{\l}{\lambda}                      
\newcommand{\r}{\rho}
\newcommand{\s}{\sigma}
\renewcommand{\t}{\tau}                         
\newcommand{\th}{\theta}
\newcommand{\f}{\phi}
\newcommand{\w}{\omega}

\begin{document}

\bibliography{journal,ashtekar,ymills,general}
\bibliographystyle{nucphys}



\begin{titlepage}
\rightline{Imperial/TP/91-92/14}
\rightline{SU-GP-91-12-2\ \ \ \ \ \ \ \ \ \ }
\begin{center}
          {\Large\bf Representations of the holonomy algebras of\\[0.2cm]
                     gravity and non-Abelian gauge theories}
\end{center}

\vspace{0.5cm}
\begin{center}
            Abhay Ashtekar${}\sp{1,2}$ and C.J.~Isham${}\sp{2}$
\end{center}
\vspace{0.5cm}
\begin{center}
                ${}\sp{1}$Physics Department\\
                Syracuse University\\
                Syracuse\\
                NY 13244-1130\\
                USA
\end{center}
\begin{center}
                ${}\sp{2}$Blackett Laboratory\\
                Imperial College\\
                South Kensington\\
                London SW7 2BZ\\
                United Kingdom
\end{center}

\begin{center}December 1991\end{center}

\begin{abstract}
Holonomy algebras arise naturally in the classical description of
Yang-Mills fields and gravity, and it has been suggested, at a heuristic
level, that they may also play an important role in a non-perturbative
treatment of the quantum theory. The aim of this paper is to provide a
mathematical basis for this proposal.

    The quantum holonomy algebra is constructed, and, in the case of real
connections, given the structure of a certain $C\sp\star$-algebra. A proper
representation theory is then provided using the Gel'fand spectral theory.
A corollory of these general results is a precise formulation of the ``loop
transform'' proposed by Rovelli and Smolin. Several explicit
representations of the holonomy algebra are constructed. The general theory
developed here implies that the domain space of quantum states can {\em
always\/} be taken to be the space of maximal ideals of the
$C\sp\star$-algebra. The structure of this space is investigated and it is
shown how observables labelled by ``strips'' arise naturally.
\end{abstract}

\end{titlepage}


\section{INTRODUCTION}
The main results of this paper are directly applicable to Yang-Mills
theory, certain topological field theories including $2+1$-dimensional
gravity, and certain systems that model $3+1$-dimensional general
relativity. However, our main motivation stems from the desire to
construct a non-perturbative quantisation of full $3+1$-dimensional
general relativity. Since this long-range goal influences several aspects
of our presentation, we will first explain our motivation and then provide
an outline of the concrete results contained in the paper.

\subsection{Motivation}
The problem of constructing a consistent (and usable) theory of quantum
gravity remains one of the major challenges facing modern theoretical
physics. Three contrasting schools of thought can be identified from the
variety of current approaches and perspectives.

    The first maintains that the construction of any consistent theory of
quantum gravity will require a profound revision of the fundamental
concepts of space, time and/or quantum theory. Iconoclastic approaches of
this kind can be exciting and thought provoking, but they suffer from the
extreme open-endedness of the problem---it is no trivial matter to rewrite
the basic laws of physics. The second, less drastic, approach is
exemplified by the current formulation of string theory. At a conceptual
level, this employs conventional quantum theoretical ideas, but the
technical framework is far removed from that of the more traditional
approaches to quantum gravity in which one tries to ``quantise'' the
standard classical theory of general relativity. In particular, the
gravitational field becomes just one among an infinity of other fields, and
is thereby deprived of much of the special status assigned to it by general
relativity.

     The third school of thought has a long history in studies of quantum
gravity. The viewpoint here is that it may well be possible to quantise
pure general relativity consistently, and in a way that respects the
geometrical framework of the classical theory, but to do so requires the
use of techniques that are quite different from the weak-field perturbative
methods that, for example, have dominated most particle-physics based
approaches to quantum gravity. Much effort has been devoted to finding such
non-perturbative schemes, and in this paper we will be concerned with a
particular one that has evolved from the introduction of a new set of
canonical variables to describe the phase space of classical general
relativity \cite{Ash87}.

    The central ingredient in this formalism is the complex $SO(3)$
connection
\footnote{A complex $SO(3)$ connection is equivalent to an $SL(2,\C)$
connection.}
\beq
A\sp{(i)}_a(x):=\Ga\sp{(i)}_a(x)+iK\sp{(i)}_a(x)\label{Def-A}
\eeq
where $\Ga$ is the usual connection on the bundle of orthonormal frames
(triads) on the spatial 3-manifold $\Si$. Thus the index $i$ ranges from 1
to 3 and can be thought of as an $SO(3)$ index, \ie $\Ga$ is a one-form on
$\Si$ taking its values in the Lie algebra of $SO(3)$. The quantity $K$ is
the usual extrinsic curvature expressed in a triadic form $K\sp{(i)}_a(x)
:= e\sp{(i)b}(x)\,K_{ab}(x)$. Here $e\sp{(i)}$, $i=1\ldots3$, is a triad
of vector fields related to the three-metric $\g$ on $\Si$ by $\g_{ab}(x)
= e\sp{(i)}_a(x)\,e\sp{(j)}_b(x)\,\de_{ij}$. As in all triadic versions of
general relativity, a new gauge invariance is present in which a triad
transforms as $e\sp{(i)}(x)\mapsto \Om\sp{i}{}_j(x)\,e\sp{(j)}(x)$ under
the action of the gauge function $\Om:\Si\map SO(3)$. The connection
property of the one-form $A$ is confirmed by its transformation as
\beq
A_a(x)\mapsto\Om(x)\,A_a(x)\,\Om(x)\sp{-1}+\Om(x)\partial_a\Om(x)\sp{-1}.
                                 \label{Trans-A}
\eeq

    A key property of the complex connection $A$ is that it forms one half
of a complete set of canonical variables, the other half being the triad
\footnote{More precisely, $e\sp{a}_{(i)}$ are vector densities of weight
$1$, but we will not need to worry about these details here.}
fields  $e\sp{a}_{(i)}$.
In particular, we have the classical Poisson bracket relations
\beqn
\{e_{(i)}\sp{a}(x),\,e_{(j)}\sp{b}(y)\}&=&0         \\
\{A\sp{(i)}_a(x),\,A\sp{(j)}_b(y)\}&=&0        \label{PB-AA}\\
\{e\sp{a}_{(i)}(x),\, A\sp{(j)}_b(y)\}&=&
            i\de\sp{a}_b\,\de\sp{j}_i\,\de\sp{(3)}(x,y).
\eeqn

    One of the most important features of the new canonical variables is
the great simplification they produce in the functional form of the
constraints. Such a simplification is highly desirable because, in
non-perturbative, Hamiltonian approaches, it is the quantum constraints
that dictate the short-distance dynamics and general behaviour of the
theory. Indeed, this is one of the main reasons for hoping that a
quantisation scheme based on these variables may yield a finite and
well-defined structure: a goal that has eluded all previous attempts to
construct a canonical theory of quantum gravity. In the present paper,
however, we are not concerned with this particular aspect of the formalism.
Our interest lies rather in another aspect, namely that use of the new
canonical variables has uncovered a close similarity between the kinematics
of general relativity and that of Yang-Mills gauge theories. This motivated
the introduction of certain quantisation techniques for gravity that are
well-adapted to deal with theories of connections but which are rather
heuristic in construction. The main purpose of this paper is to give a
proper mathematical meaning to the kinematical features of these
quantisation schemes.

    For example, (\ref{PB-AA}) implies that the associated quantum fields
$\op{A}\sp{(i)}_a(x)$ commute, which suggests that these fields might be
simultaneously diagonalised in a representation \cite{Ash86,JS88} in which
the state vectors are functionals $\Psi$ on the space $\A$ of all
connections, and where $\big(\op{A}\sp{(i)}_a(x)\Psi\big)(A)$ $:=$
$A\sp{(i)}_a(x)\Psi(A)$. The gauge invariance under local triad rotations
can then be imposed in various ways. One possibility is to fix the gauge
before quantising; another is to attempt to use gauge-invariant observables
from the outset. This second possibility was developed in the important
work of Rovelli and Smolin \cite{RS90,Rov91} based on a new set of
variables which exploit the fact that $A$ is a connection. These are
defined by
\beqn
T_\a(A,e) &:=& \half\Tr\left({\cal P}\exp\oint_\a A\right)
                                                \label{Def-T0}\\
T\sp{a}_{\a,s}(A,e) &:=& \Tr\left({\cal P}
            e\sp{a}(\a(s))\exp\oint_\a A\right) \label{Def-T1}
\eeqn
where $\cal P$ denotes the path-ordered line integral around the loop $\a$
in $\Si$.

    These new, non-local, variables have a closed Poisson bracket algebra,
and a central assumption is that there exist genuine well-defined operator
equivalents in the quantum theory. Such a requirement is very non-trivial
since, for example, in the case of the $T$-variables in (\ref{Def-T0}) it
implies that the underlying operator field $\op{A}$ can be smeared with a
Dirac $\de$-function concentrated on a loop, rather than with one of the
smooth test functions of conventional quantum field theory. This is
certainly not possible for the standard Fock space operators and, indeed,
specific examples \cite{AR91,Ash91} of Maxwell theory and $2+1$ gravity
show that considerable care is needed
\footnote{See also our recent work \cite{AI92}.}.

     Rovelli and Smolin used the formal representation of the algebra
generated by $\hat{A}_a\sp{(i)}$ and $\hat{e}_{(i)}\sp{a}$ in which the
state vectors are functionals of gauge-equivalence classes $[A]_\G$ of
connections, and argued that the operators corresponding to
(\ref{Def-T0}-\ref{Def-T1}) can be represented on these states. For
example, the $T$-operators of (\ref{Def-T0}) act as
\beq
\big(\op T_\a\Psi\big)([A]_\G):=\half\Tr\left({\cal P}\exp\oint_\a
                A\right)\Psi([A]_\G).       \label{Def-RepT0}
\eeq
However, no inner products were specified. This is not surprising since it
is a highly non-trivial mathematical problem (and one that arises already
in conventional Yang-Mills theory) to see if such functionals can be
equipped with a Hilbert space structure involving a measure on the space
$\A/\G$ of orbits of the gauge group $\G$ acting on $\A$. For example, in
the case of a field theory whose configuration space is some
infinite-dimensional topological vector space $E$, it is well known that
the state vectors (and measures) have usually to be defined on the
topological {\em dual\/} space $E'$ of distributions \cite{GV64,Yam85}. It
might be anticipated that something similar happens in the present case.
However, $\A/\G$ has a complicated topological structure and is certainly
not a vector space. So what is meant by a ``distributional'' analogue of an
element of $\A/\G$? This intriguing question is not just of mathematical
interest: it also has a direct physical relevance since, as in all quantum
theories, it is the inner product on the state space that yields the
probabilistic predictions of the theory. This vital issue of the inner
product has received scant attention in the literature so far and
constitutes one of the main motivations for our investigations.

     Rovelli and Smolin also constructed a ``loop representation'' (in
which states are functionals of loops) of the algebra and suggested that
the connection and loop representations can be related by a ``loop
transform''
\beq
\twid{\Psi}(\a) := \int_{\A/\G}{\cal D}([A]_\G)\,\Tr\left({\cal P}
\exp\oint_\a A\right)\Psi([A]_\G).              \label{Formal-LT}
\eeq
Once again, we see the need to construct measures on the non-linear space
$\A/\G$ (or its ``distributional'' dual).

\subsection{Outline of the paper}
After these motivational remarks we can now summarise the major problems
addressed in the present paper.
\be
\item We wish to discuss the construction of inner products on the vector
spaces of states. In particular, we are interested in the possibility of
defining measures $\mu$ on $\A/\G$ in order to form the Hilbert spaces
$L\sp{2}(\A/\G,d\mu)$.

\item This raises the general questions of
    \bi
    \item is it necessary to introduce ``distributional'' analogues of
elements of $\A/\G$?
    \item what might be meant by such objects?
    \ei
\item  A related problem is to give a proper meaning to the loop transform
defined heuristically in (\ref{Formal-LT}). In particular, which is more
fundamental: the connection representation $\Psi([A]_\G)$ or the loop
representation $\twid{\Psi}(\a)$? And what is the precise relation between
them?
\ee
To address these issues, we proceed as follows. We begin by considering
the abelian quantum algebra (the ``holonomy algebra'') generated by the
operator version of the $T_\alpha$ variables defined in (\ref{Def-T0}).
The crucial idea is to endow it with the structure of a
$C\sp\star$-algebra and then use the Gel'fand spectral theory to
systematically analyse its cyclic representations. The key results that we
need from the Gel'fand theory are the following:
\be
\item The space $\D$ of maximal ideals of a commutative $C\sp\star$-algebra
is compact (and Hausdorff) in a natural topology.
\item The given $C\sp\star$-algebra is isomorphic to $C(\D)$, the algebra
of continuous functions on $\D$.
\ee
By applying these results to the holonomy $C\sp\star$-algebra we show
that, in any of its cyclic representations, the Hilbert space of states
can be identified with $L\sp{2}(\D,d\mu)$ for some regular measure $\mu$
on $\D$, and that the holonomy operators act simply by multiplication. In
particular, these results enable us to give a precise meaning to the formal
representation (\ref{Def-RepT0}) and the Rovelli-Smolin transform
(\ref{Formal-LT}). Finally, we explore the structure of the space $\D$
which is now the ``domain space'' for quantum states. Not surprisingly,
$\A/\G$ appears as a proper subspace of $\D$. We identify the elements of
$\D-\A/G$ as the ``distributional analogues'' of gauge-equivalence classes
of smooth connections and present examples of such entities.

     In order to endow a $C\sp\star$-structure on the holonomy algebra, we
are obliged to make a drastic simplifying assumption: namely, that the
connection $A_a\sp{(i)}$ is {\em real\/}. Consequently, the main results
of the present paper are applicable only to certain model systems and not
to the full, $3+1$-dimensional general relativity. However, the simpler
systems which do satisfy this assumption are of considerable interest in
their own right. They include the genuine $SU(2)$ Yang-Mills theory,
several of the topological quantum field theories, the Husain-\Kuchar
\cite{HuK90} model for gravity, $2+1$-dimensional general relativity, and
the midi-superspace of $3+1$-dimensional gravitational fields with one
Killing vector.

    The Hamiltonian structure of the Husain-\Kuchar model is essentially
the same as that of $3+1$-dimensional general relativity {\em without\/},
however, the scalar (or, the Hamiltonian) constraint. The configuration
variable in this model is a real, $SU(2)$-valued connection. In its
connection-dynamics version, $2+1$-dimensional gravity is structurally very
similar to the $3+1$-dimensional theory in the new canonical variables
\cite{Wit88,AHRSS89}. However, in the $2+1$-dimensional case, the
connection is real and flat and takes its values in the Lie algebra of
$SO(2,1)$. Finally, in the presence of a space-like Killing vector,
$3+1$-dimensional general relativity is equivalent to $2+1$-dimensional
gravity coupled to a doublet of scalar fields (which constitute a
non-linear $\s$ model). The connection is now only a part of the complete
set of configuration variables. However, it is {\em real\/}, generally
non-flat, and $SO(2,1)$-valued. Even though this last example goes a long
way towards full, $3+1$-dimensional general relativity, the methods
developed in this paper cannot be applied directly to the full theory where
the connection is genuinely complex. The reality assumption on
$A_a\sp{(i)}$ is therefore a severe limitation and one to which we hope to
return in a later work.

     The paper is organised as follows. In section \ref{Sec:Algebra}, we
construct the holonomy $C\sp\star$-algebra and analyse the structure of
its cyclic representations. We then illustrate the underlying ideas using
explicit, albeit simple, examples. In particular, we are able to give a
precise mathematical meaning to the Rovelli-Smolin ``loop transform'' in a
general setting and reveal the structure involved via examples. Section
\ref{Sec:Ideal-Space} is devoted to the study of the space $\D$ of maximal
ideals of the holonomy algebra, on which states and integrals are defined,
with an emphasis on its distributional elements. To find concrete examples
of these entities, we first construct a large class of distributional,
$\G$-equivariant group transformations on the space $\A$ that can be
viewed as the vector-field analogues of the Rovelli-Smolin ``momentum
variables'' defined in (\ref{Def-T1}). We then show how families of such
transformations, when integrated over 2-dimensional strips in $\Si$, can
produce distributional elements of $\D$. The paper concludes with a list
of open issues whose resolution would greatly clarify the structure of
$\D$ and make the task of finding physically interesting representations
of the holonomy algebra more manageable.


\section{REPRESENTATIONS OF THE ABELIAN T-ALGEBRA}
\label{Sec:Algebra}
This section is divided into four parts. In the first, we recall some basic
ideas concerning holonomies of $SL(2,\C)$ connections; in the second we
construct the $C\sp\star$-holonomy algebra; in the third we discuss its
representations and provide a precise mathematical meaning to the
Rovelli-Smolin loop transform; and, in the fourth, we present a few simple
examples to illustrate the constructions involved.

\subsection{Preliminaries}\label{Sec:IIA}
The main aim of this section is to study the general representations of the
commutative operator algebra generated by the classical variables defined
in (\ref{Def-T0}). To do this properly we must start by defining these
objects more carefully. First we choose some fixed base-point $x_0$ in the
three-manifold $\Si$ and let $\Lx$ denote the set of all continuous,
piecewise smooth loops on $\Si$ which map
\footnote{The points $0$ and $2\pi$ are identified in the parametrisation
of the points on the circle $S\sp{1}$ by the real numbers in the closed
interval $[0,2\pi]$.}
the base-point $0\in[0,2\pi]\simeq S\sp{1}$ to $x_0\in\Si$. We are
interested only in parametrised loops \ie we identify the loops
$\a:S\sp{1}\map\Si$ and $\a\c\f:S\sp{1}\map\Si$ for all
orientation-preserving diffeomorphisms $\f:S\sp{1}\map S\sp{1}$ which map
the base-point $0\in S\sp{1}$ to itself.

    It is important to note that $\Lx$ is a semigroup under the combination
law
\footnote{It is conventional to use $\a\c\b$ to denote the combination of a
pair of loops. However, it should be noted that the symbol $\c$ does {\em
not\/} then have its usual meaning of the composition of maps.}
$\a,\b\mapsto\a\c\b$ where $\a\c\b$ means the loop formed by going first
around $\b$ and then around $\a$, and where the identity element is defined
to be the constant loop $x_0(t):=x_0,\forall t\in S\sp{1}$. This semigroup
structure is compatible with the equivalence relation that identifies loops
which differ by a reparametrisation map $\f:S\sp{1}\map S\sp{1}$. However,
we do not get a {\em group\/} structure in this way since, not withstanding
the notation (and occasional implications in the literature to the
contrary), the inverse loop $\a\sp{-1}$ defined by
$\a\sp{-1}(t):=\a(2\pi-t)$ is not a group inverse, \ie $\a\c\a\sp{-1}\ne
x_0$. On the other hand, the inverse loop operation is a genuine {\em
involution\/} on the semigroup in the sense that, for all $\a,\b\in\Lx$,
$(\a\c\b)\sp{-1}=\b\sp{-1}\c\a\sp{-1}$, and, $(\a\sp{-1})\sp{-1} = \a$.

    The next step is to consider the parallel transport around such loops
associated with a connection. Although the general ideas of the paper are
applicable to any gauge group, because of our interest in gravity, the most
interesting case for us is $SL(2,\C)$ and subgroups thereof. For
definiteness, let us therefore consider a smooth connection in an arbitrary
principle $SL(2,\C)$ bundle over $\Si$. We will denote by $\A$ the set of
all such connections and by $\G$ the gauge group of automorphisms of the
bundle
\footnote{We assume that $\A$ is one of the standard function spaces. If
$\Si$ is non-compact it is necessary to impose asymptotic fall-off
conditions on both $\A$ and $\G$ in order that, for example, $\A/\G$ be a
genuine infinite-dimensional manifold \cite{MV81}.}.

    A connection should really be viewed as a one-form on the bundle space
of the bundle. However, since all $SL(2,\C)$ (and $SU(2)$) bundles over a
three-manifold are necessarily trivial, there always exists a global
section (in the case of the bundle of frames, this is a global triad) that
can be used to ``pull-down'' the connection one-form to give a one-form $A$
on $\Si$. The parallel transport around a loop $\a\in\Lx$ is a map from the
fiber over $x_0\in\Si$ to itself, and can only be identified with an
element of the structure group once a base point in this fiber has been
chosen. In our case, the global section can be used for this purpose and,
with this proviso, it is meaningful (and correct) to write the parallel
transport in the familiar way as the group element $P_\a(A)$ defined by
\beq
    P_\a(A):= {\cal P}\exp\oint_\a A    \label{Def-Pa}
\eeq
which belongs to $SL(2,\C)$ (or, in the real case, $SU(2)$). The
$T$-variables are then defined on the classical phase space of all pairs
$(A,e)$ as in (\ref{Def-T0}):
\beq
   T_\a(A,e) := \half\Tr\big(P_\a(A)\big).  \label{Def-T0'}
\eeq

    At this point it should be re-emphasised that the physical
configuration space of the system is really the quotient space $\A/\G$
which, with care, \cite{MV81} can be given the structure of an
infinite-dimensional, topologically non-trivial, manifold. An important
step in the development of a general quantisation scheme \cite{Ish84} is
the construction of an (over-) complete set of functions on $\A/\G$. It is
therefore significant that each $T_\a$ is a gauge-invariant functional of
$A$ and, as such, projects down to give a functional on $\A/\G$. In the
$SU(2)$ case, these functionals form a separating set on $\A/\G$: if all
the $T_\a$ assume the same values at two connections, they are necessarily
gauge related (this is proved in detail in section \ref{Sec:Ideal-Space}).
In the case when the group is $SL(2,\C)$ this result continues to hold
\cite{GLS91} except for a well-defined set ``of measure zero''.
Furthermore, it is often asserted that {\em all\/} gauge-invariant
functionals on $\A$ can be written in terms of these Wilson-loop
functionals, although it is difficult to find an exposition of the precise
meaning of ``all'' and ``can be written''. We will return to this issue
later in section \ref{Sec:Ideal-Space}.

    From now on we will view the $T$-variables as functions on the physical
configuration space $\A/\G$, and begin to consider the implications of
their use in the quantum theory. The vanishing of the classical Poisson
bracket of $A\sp{(i)}_a(x)$ with $A\sp{(j)}_b(y)$ suggests that the
corresponding quantum operators should commute and hence that
\beq
    \[\op T_\a,\op T_\b\]=0.                \label{Com-TaTb}
\eeq
for all $\a,\b\in\Lx$. This in turn suggests that we might start by
considering the spectral theory of the abelian algebra of the operators in
(\ref{Com-TaTb}). However, we must first address the crucial fact that the
classical $T$-variables are not functionally independent but instead
satisfy a well-known set of identities which have their origin in the
relations
\beq
   P_\a(A)\,P_\b(A) = P_{\a\c\b}(A)     \label{PaPb=}
\eeq
for all $\a,\b\in\Lx$ and $A\in\A$. Since $P_{x_0}=\unit$ this means that,
for each $A\in\A$, the map $\a\mapsto P_\a(A)$ provides a matrix
representation of the semigroup $\Lx$ in which $P_{\a\sp{-1}}(A) =
(P_\a(A))\sp{-1}$. However, any pair of $2\times2$ unit determinant
matrices $C,D$ satisfy the identity
\beq
  \Tr C\,\Tr D \equiv \Tr(CD) + \Tr(CD\sp{-1})
\eeq
and, in addition, any unit determinant $2\times2$ matrix $C$ satisfies
\beq
  \Tr C \equiv \Tr C\sp{-1}.
\eeq
These identities imply in particular that the $T$-variables satisfy the
non-linear relations
\beqn
T_\a T_\b &=& \half(T_{\a\c\b} + T_{\a\c\b\sp{-1}}) \label{Man1}\\
T_\a &=& T_{\a\sp{-1}}                              \label{Man2}
\eeqn
which are nothing other than the famous Mandelstam relations \cite{Man79}
for the special case of $SL(2,\C)$. That there exist such algebraic
relations between $T_\a$ is not surprising: since, $\A/\G$ is a genuine
manifold with a complicated topology, {\em any\/} set of globally-defined
functions that separates points of $\A/\G$ is necessarily overcomplete. The
crucial question is how these relations are to be incorporated in the
quantum theory.

    One possibility is to greatly reduce the number of $T_\a$-variables by
solving as many algebraic constraints as possible at the classical level
{\em before\/} quantising. For example, this approach has been adopted
recently by Loll in the context of a lattice approximation to the theory
\cite{Lol91}. The other possibility, which we will explore here, is to
assume that there exist genuine operator analogues $\op{T}_\a$ of the
classical $T$-variables, and then to impose the constraints on them. This
could be done in various ways, and we conclude this sub-section by
discussing one possibility, based on the theory of representations of
groups, that has been used in the literature. However, note that the avenue
we will actually follow in this paper is somewhat different. It employs the
representation theory of $C\sp\star$-algebras and will be discussed in
detail in the next sub-section.

    A first impulse may be to start with (\ref{PaPb=}) and explore the
general theory of the representations of the semigroup $\Lx$. A priori,
this is not too easy since this semigroup is not abelian, but the form of
the classical $T$-variables suggests a modified strategy. Let us define the
equivalence relation (to be denoted $R$) on $\Lx$
\beq
 \a\equiv\b {\rm\ mod\ } R{\rm\ if\ }T_\a(A)=T_\b(A){\rm\ for\ all\ }
 A\in\A
\eeq
and associate operators only with the equivalence classes $[\a]_R$
(following the fact that, by the definition of the $R$ equivalence
relation, the classical $T$-variables depend only on the equivalence
classes of loops). Note that the equivalence relation implies:
\beqn
    \a\c\b &\equiv&\b\c\a.               \label{PropR1}\\
    \a &\equiv&\a\sp{-1}                 \\
    \a\c\a\sp{-1}&\equiv& x_0           \\
    \a\c\r\c\r\sp{-1}&\equiv&\a         \label{PropR4}
\eeqn
for all $\a,\b\in\Lx$, where (\ref{PropR4}) applies for any path $\r$ in
$\Si$ which has an end that touches the loop $\a$ somewhere.

    Now, it is tempting to try to define a product on the set $\L_{x_0}/R$
of equivalence classes by
\beq
  [\a]_R\,[\b]_R := [\a\c\b]_R            \label{RaRb=?}
\eeq
since, by virtue of (\ref{PropR1}), this would yield an abelian structure.
However, the meaning of $\a_1\equiv\a_2$ is that, for all $A\in\A$, $\Tr
P_{\a_1}(A)$ = $\Tr P_{\a_2}(A)$, and this does {\em not\/} imply that, for
all $\b\in\Lx$, $\Tr\big(P_{\a_1}(A)\,P_\b(A)\big)$ =
$\Tr\big(P_{\a_2}(A)\,P_\b(A)\big)$, which has to be the case if the
definition (\ref{RaRb=?}) is to be consistent. (It works if $\a_2 =
\a_1\c\r\c\r\sp{-1}$ but, for example, not if $\a_2=\a_1\sp{-1}$.)

    One way of overcoming this problem is first to define a loop $\g$ to be
{\em thin\/} \cite{GT80,Ana83,Bar91} if there exists a homotopy of $\g$ to
the trivial loop in which the image of the homotopy lies within the image
of $\g$. Then a new relation can be introduced on $\Lx$ by saying that two
loops $\a$ and $\b$ are {\em thinly equivalent\/} (denoted $\a\equiv\b{\rm\
mod\ }t$) if $\a\c\b\sp{-1}$ is a thin loop. Note that $\a\equiv\b$ mod $t$
implies $\a\equiv\b$ mod $R$ but not conversely. In particular,
$\a\equiv\a\sp{-1}$ mod $R$ whereas $\a\c\a$ is generally not thin. Thus
there exists a non-injective surjection $\Lx/t\map\Lx/R$ in which
$[\a]_t\mapsto[\a]_R$. It is easy to show that a well-defined composition
law on $\Lx/t$ is given by $[\a]_t\,[\b]_t :=[\a\c\b]_t$. This gives an
abelian group whose representations can be studied using standard
techniques. A similar path has been followed in the past \cite{Gil81,GT86}
with some success. However, in order to address the particular problems in
which we are interested it is advantageous to proceed in a somewhat
different way which we discuss in detail in the next sub-section.
Nonetheless, the mathematical structure sketched above---especially the
equivalence relations---will turn out to be useful at several points in the
development of our approach.

\subsection{The $C\sp\star$-algebra for the $T$-variables}\label{Sec:IIB}
In the classical theory, the functions $T_\a$ on $\A/\G$ generate an
abelian, associative algebra. In this sub-section, we wish to introduce an
analogous algebra of quantum operators $\op{T}_\a$. Thus, we wish to define
quantum operators $\op{T}_\a$ such that
\beq
 \op{T}_{\a}\,\op{T}_{\b}=\half\left(\op{T}_{\a\c\b}+
        \op{T}_{\a\c\b\sp{-1}}\right)            \label{Op-Man1}
\eeq
or, more precisely,
\beq
 \op{T}_{[\a]_R}\,\op{T}_{[\b]_R}=\half\left(\op{T}_{[\a\c\b]_R}+
 \op{T}_{[\a\c\b\sp{-1}]_R}\right).             \label{Op-TaTb}
\eeq
Once again, it should be emphasised that even the assumption that the
$\op{T}_\a$ operators {\em exist\/} is a radical departure from
conventional quantum field theory.

    The crucial question is how best to study the general structure of such
operators. Our approach is to construct a $C\sp\star$-algebra in which the
multiplication law is defined in such a way that any representation of the
algebra is guaranteed to produce operators satisfying (\ref{Op-TaTb}). To
this end let us start by defining $\HA$ to be the set of all finite complex
linear combinations of the classical $T$-variables. The key observation is
that, by virtue of the classical Mandelstam identities (\ref{Man1}), this
vector space of functions on $\A/\G$ is {\em closed\/} under the usual
product law of functions. We will exploit this feature in our construction
of the quantum algebra. Note how important it is that the structure group
is $SL(2,\C)$ or a sub-group thereof: for $SU(n)$, $n\ge 3$, the Mandelstam
identities involve products of {\em three\/} or more $T_\a$ variables and
so the set $\HA$ is no longer closed under multiplication. This is one of
those relatively rare cases where there is a significant difference between
$SL(2,\C)$ and any other internal symmetry group
\footnote{For the cases $n\ge 3$ it would be possible to use the algebra
{\em generated\/} by products of the $T_\a$ variables. However, several of
the constructions that follow would be significantly more complicated in
this case.}.

    To construct the required quantum algebra, it is convenient to extract
the essential algebraic features of $\HA$. To do so, let us first consider
the vector space $F\Lx$ of all finite, complex linear combinations of
elements of $\Lx$. Equivalently, this is the set of all complex-valued
functions on $\Lx$ whose supports are finite subsets of $\Lx$. The sum is
defined pointwise in the usual way, as is the product of any finite sum by
a complex number. The crucial thing is the product between a pair of finite
sums $\sum_{i=1}\sp{n}a_i\a_i$ and $\sum_{j=1}\sp{m}b_j\a_j$. This is
defined in the standard way as
\beq
 \left(\sum_{i=1}\sp{n}a_i\a_i\right)\left(\sum_{j=1}\sp{m}b_j\a_j\right)
   := \sum_{i=1}\sp{n}\sum_{j=1}\sp{m}a_ib_j\a_i\a_j
                            \label{Def-prod1}
\eeq
where $n$ and $m$ are any finite integers, and $a_i,b_j\in\C$. Hence the
key step is to define the product in $F\Lx$ of a pair of loops
$\a,\b\in\Lx$. We choose
\beq
   \a\b:=\half(\a\c\b+\a\c\b\sp{-1})          \label{Def-prod2}
\eeq
which builds the $SU(2)$ (or $SL(2,\C)$) Mandelstam identity into the very
fabric of the algebra.

    Any operator representation of this algebra will necessarily satisfy
the relations (\ref{Op-Man1}). Recall, however, from section
(\ref{Sec:IIA}) that the classical functions $T_\a$ depend not on specific
loops $\a$ but only on equivalence classes, $[\a]_R$. As matters stand,
there is no a priori reason why any specific representation of our quantum
algebra should respect the extra requirement that the operators $\op{T}_\a$
depend only on the $R$-equivalence classes of the loops. In addition, it is
hard to exploit the algebra directly since it is both non-abelian and
non-associative. Finally, the constant loop $x_0$ is not a unit for this
algebra since $x_0\b$ = $\half(\b+\b\sp{-1})$, which does not equal $\b$
because $\b\ne\b\sp{-1}$. These problems can all be resolved by
incorporating more of the properties of the classical $T$-variables.
Specifically we consider the linear subspace $K$ of $F\Lx$ defined by
\beq
 K := \big\{x=\sum_{i=1}\sp{n}a_i\a_i\big|\sum_{i=1}\sp{n}a_iT_{\a_i}(A)=0
 {\rm\ for\ all\ } A\in\A \big\}          \label{Def-K}
\eeq
which has the important property of being a two-sided {\em ideal\/} in the
algebra $F\Lx$. For example, if $\sum_{i=1}\sp{n}a_i\a_i\in K$ then
$(\sum_{i=1}\sp{n}a_i\a_i)\b$ =
$\half\sum_{i=1}\sp{n}a_i(\a_i\c\b+\a_i\c\b\sp{-1})$. However,
$\half\sum_{i=1}\sp{n}a_i(T_{\a_i\c\b}+T_{\a_i\c\b\sp{-1}})$ =
$(\sum_{i=1}\sp{n}a_iT_{\a_i})T_\b=0$ and so $(\sum_{i=1}\sp{n}a_i\a_i)\b$
also belongs to $K$. It follows at once that $(\sum_{i=1}\sp{n}a_i\a_i)x\in
K$ for all $x\in F\Lx$, and there is a similar proof for left
multiplication by $x$. Since $K$ is an ideal, a product can be defined
consistently on the quotient space $F\Lx/K$ by $[x]_K\,[y]_K:=[xy]_K$ where
$[x]_K$ denotes the $K$-equivalence class of $x\in F\Lx$.

    Note that a natural map $F\Lx\map\HA$ is given by
$\sum_{i=1}\sp{n}a_i\a_i$ $\mapsto$ $\sum_{i=1}\sp{n}a_iT_{\a_i}$ and it is
clear that the kernel is precisely $K$. Hence $F\Lx/K$ is isomorphic to the
holonomy algebra $\HA$
\footnote{The reason we did not just begin with $\HA$ but arrived at it
starting from $F\Lx$ is that the explicit construction brings out the
algebraic structures involved. This will be useful in the next section in
our discussion of representations of this algebra.}.
{}From now on, we will denote the generators of this algebra $F\Lx/K\simeq
\HA$ either by $[\a]_K$, or, when there is no danger of confusion, simply
by $[\a]$. The algebra has the following important properties:
\be
\item It is associative and abelian.

\item It has a unit element $e:=[x_0]_K$.

\item If $\a\equiv\b$ mod $R$, then $T_\a=T_\b$ and hence $\a-\b\in K$.
Thus any operator representation of $F\Lx/K$ has the desired property that
$\op{T}_\a$ depends only on the $R$-equivalence class of $\a\in\Lx$.

\item Any representation of this algebra necessarily satisfies the operator
Mandelstam identities
\footnote{It also satisfies the operator equivalent of the Mandelstam
identity (\ref{Man2}); \ie $\op{T}_\a=\op{T}_{\a\sp{-1}}$.}
(\ref{Op-Man1}).
\ee
\noindent Note also that there is a natural isomorphism of $F\Lx/K$ with
$F{\cal L}_{x_0'}/K$ where $x_0'$ is any other choice of base point. This
is obtained by connecting $x_0'$ to $x_0$ with any smooth path $\eta$ and
then noting that the trace of the parallel transport along the curve
$\eta\sp{-1}\c\a\c\eta$ (a loop based at $x_0'$) is equal to the trace of
the transport around the loop $\a$ based at $x_0$. Thus, although it is
convenient to have a fixed base point $x_0$ at one's disposal, the
structure of the holonomy algebra itself is {\em independent\/} of the
choice of this point.

    The next step is to convert $F\Lx/K$ into a proper normed
$\star$-algebra to which the powerful tools of spectral theory can be
applied. This can be done in several ways. If the gauge group is $SU(2)$,
the most obvious approach is to exploit the fact that each classical $T_\a$
is a {\em bounded\/} function on $\A/\G$. This follows since any $SU(2)$
matrix $M$ can be written in the form
$M=\left({a\atop -\bar{b}}{b\atop\bar{a}}\right)$ where
$|a|\sp{2}+|b|\sp{2}=1$. Thus $|\Tr M|$ = $|a+\bar{a}|\le 2|a|\le 2$ which
shows that $|T_\a(A)|\le 1$ for all $\a\in\Lx$ and $A\in\A$. Thus we can
write $\HA\subset B(\A/\G)$ where $B(\A/\G)$ denotes the set of all
complex-valued, bounded functions on $\A/\G$. It follows that a natural
norm on $F\Lx/K$ is
\beq
 \big\|\sum_{i=1}\sp{n}a_i[\a_i]_K\big\| :=
 \sup_{[A]_\G\in\A/\G}\big|\sum_{i=1}\sp{n}a_i T_{\a_i}([A]_\G)\big|
                                      \label{Def-norm}
\eeq
and it is a trivial piece of analysis to show that, for all $x,y\in
F\Lx/K$, $\norm{xy}\le\norm{x}\,\norm{y}$ so that the algebra
multiplication is continuous with respect to the norm topology on $\HA$.
However, this strategy does {\em not\/} work when the group is $SL(2,\C)$
since the trace function on these matrices is not bounded; one must use a
different avenue to induce a topology on $\HA$. We will see at the end of
section \ref{Sec:IIC} that the general framework is rather insensitive to
the precise choice of topology. However, we will not discuss the available
choices, since we are about to restrict ourselves to $SU(2)$ for a quite
different reason.

    This step is occasioned by a major problem that arises in the
$SL(2,\C)$-case when we try to convert $F\Lx/K\simeq\HA$ into a
$\star$-algebra. The difficulty occurs already at the classical level when
we enquire into the reality of the $T$ variables defined in
(\ref{Def-T0'}). For the $SU(2)$ (or, $SU(1,1)$) case there is no problem
since these variables are automatically real. This follows at once from the
representation of any $SU(2)$ matrix $M$ as
$M=\left({a\atop -\bar{b}}{b\atop\bar{a}}\right)$ with
$|a|\sp{2}+|b|\sp{2}= 1$, (or, of a $SU(1,1)$ matrix as
$\left({a\atop\bar{b}}{b\atop\bar{a}}\right)$, with $|a|\sp{2}-|b|\sp{2} =
1$), whose trace $a+\bar{a}$ is always real. This result can be coded into
the Banach algebra by defining the adjoint operation as
\beq
 \Big(\sum_{i=1}\sp{n}a_i\,[\a_i]_K\Big)\sp\star
   :=\sum_{i=1}\sp{n}\bar{a_i}\,[\a_i]_K            \label{Def-star}
\eeq
so that, for any loop $\a\in\Lx$, the $K$-equivalence class $[\a]_K$ is a
self-adjoint element of the algebra.

    When equipped with the $\star$-operation of (\ref{Def-star}), $\HA$
becomes a commutative $\star$-algebra whose completion with respect to the
norm (\ref{Def-norm}) is a $C\sp\star$-algebra which will be denoted
$C\sp\star(\HA)$. This is the required holonomy $C\sp\star$-algebra in the
$SU(2)$ (or $SU(1,1)$) case. The difficulty with the general $SL(2,\C)$
case is that, even at the classical level, there is no apparent way in
which the complex conjugate $\bar{T_\a}$ can be expressed as a linear
combination of the $T$ variables. Thus, in this case, the algebra $\HA$ is
not closed under complex conjugation. This is therefore the point at which
we must make our restrictive assumption that the connection---and hence
$T_\a$---is real. For definiteness, in the main body of this paper, we will
consider the $SU(2)$-case and defer discussion of the $SU(1,1)$ theory to a
later work.

    The $C\sp\star$-algebra $C\sp\star(\HA)$ is of considerable importance
for the following reason. We are interested in representations of the
algebra $\HA$. Let $\op{R}:\HA\map B(\H)$ be any continuous representation
of $\HA$ with bounded operators on some Hilbert space $\H$. Thus
\be
\item $\op{R}(x+y)=\op{R}(x)+\op{R}(y)$ for all $x,y\in\HA$;

\item $\op{R}(xy)=\op{R}(x)\op{R}(y)$ for all $x,y\in\HA$;

\item $\op{R}(x\sp*)=\op{R}(x)\sp{\dagger}$ for all $x\in \HA$.
\ee
\noindent Then $\op{R}$ can be extended to the complete algebra
$C\sp\star(\HA)$ since $\HA$ is a dense subset. Conversely, any
representation of $C\sp\star(\HA)$ clearly passes to a representation of
the subset $\HA$. Hence the continuous representations of $\HA$ are in
one-to-one correspondence with those of the abelian $C\sp\star$-algebra
$C\sp\star(\HA)$.

    Representations of $C\sp\star$-algebras can be analysed systematically
using the powerful machinery of Gel'fand spectral theory. The key result
here is that {\em any commutative\/} $C\sp\star$-{\em algebra} $\Alg$ {\em
is isomorphic to the algebra\/} $C(\D)$ {\em of all continuous,
complex-valued functions on the space\/} $\D$ {\em of maximal ideals in\/}
$\Alg$. To specify the isomorphism, let us first recall that there is a
one-to-one correspondence between maximal ideals and linear, multiplicative
homomorphisms $h$ from $\Alg$ to $\C$ (\ie $h(xy)=h(x)\,h(y)$ for all
$x,y\in\Alg$; for a $C\sp\star$-algebra any such homomorphism also
satisfies $h(x\sp\star)=\bar{h(x)}$ for all $x\in\Alg$). The Gel'fand
transform which implements the isomorphism between $\Alg$ and $C(\D)$ is
the map defined by
\beqn
      \Alg&\map &C(\D)    \label{Gel-trans}\\
    x&\mapsto&\twid{x}               \nonumber
\eeqn
where $\twid{x}(h):=h(x)$ for all multiplicative homomorphisms
$h:\Alg\map\C$.

    The ideal space $\D$ is given the Gel'fand spectral topology, which is
the weakest topology such that all the functions of the form
$\twid{x}:\D\map\C$, $x\in\Alg$, are continuous. A key result is that {\em
in this topology,} $\D$ {\em is a compact, Hausdorff space}. It therefore
admits regular measures with respect to which its total volume is finite.
This suggests that it may be possible to represent quantum states by
functions on $\D$, and with the inner product defined via integrals on this
space. We will see that this expectation is correct. We will also see that,
for the holonomy $C\sp\star$-algebra under consideration, the classical
configuration space $\A/\G$ is in fact naturally embedded in $\D$. The
precise structure of $\D$ is clearly of considerable importance to the task
of understanding the quantum theory, and much of the discussion in section
\ref{Sec:Ideal-Space} is devoted to this problem.

    We conclude this sub-section by noting certain properties of $\D$ which
will be useful in the rest of this section. Any $h$ in the maximal ideal
space $\D$ of $C\sp\star(\HA)$ defines an associated map (also denoted $h$)
from $\Lx$ to $\R$ which satisfies the three conditions
\beq
h(\a)h(\b)=\half\big(h(\a\c\b)+h(\a\c\b\sp{-1})\big), \label{Prop-h1}
\eeq
\beq
  {\rm if\ } \sum_{i=1}\sp{n}a_i\a_i\in K{\rm\ then\ }
     \sum_{i=1}\sp{n}a_ih(\a_i)=0,              \label{Prop-h2}
\eeq
\beq
            |h(\a)|\le 1.                       \label{Prop-h3}
\eeq
These follow simply from the properties of the algebra $C\sp\star(\HA)$ and
the fact that $h$ is a homomorphism. There is a partial converse to the
above in the sense that any function $h:\Lx\map\R$ which satisfies
(\ref{Prop-h1}--\ref{Prop-h2}) extends by linearity to give a function
$h:F\Lx/K\simeq\HA\map\C$ which is a multiplicative homomorphism on the
subalgebra $\HA$ of $C\sp\star(\HA)$. However, whether or not this extends
to a homomorphism on $C\sp\star(\HA)$ itself depends on the detailed
behaviour of $h$ on infinite sets of loops. In particular, imposing the
boundedness condition (\ref{Prop-h3}) is not obviously sufficient to
guarantee such an extension when $\HA$ is completed with respect to the
sup-norm topology. This is one area in which it could be useful to try to
find a slightly different, and more convenient, topology on $\HA$.

\subsection{The loop transform and representations of the holonomy
algebra}\label{Sec:IIC}
The results above can now be used to give a precise meaning to the
representation and loop transform defined heuristically in
(\ref{Def-RepT0}) and (\ref{Formal-LT}) respectively. Let
$\op{R}:C\sp\star(\HA)\map B(\H)$ be {\em any\/} continuous representation
of $C\sp\star(\HA)$ on a Hilbert space $\H$ with a cyclic vector $\Om$. The
function $\Ga(x):=\la\Om,\op{R}(x)\Om\ra$ defines a positive linear
functional on $C\sp\star(\HA)$. However, $C\sp\star(\HA)$ is isomorphic to
$C(\D)$ via the Gel'fand transform $x\mapsto\twid{x}$, and hence we get a
continuous, positive-definite linear function $\twid{\Ga}$ on $C(\D)$
defined by
\beq
   \twid{\Ga}(\twid{x}):=\Ga(x).
\eeq
Since $\D$ is a compact space, the Riesz lemma shows that there exists some
regular measure $\mu$ on $\D$ such that
\beq
  \la\Om,\op{R}(x)\Om\ra = \int_{\D}d\mu(h)\,\twid{x}(h)
        =\int_{\D}d\mu(h)\,h(x)
\eeq
and, in particular,
\beq
  \la\Om,\op{T}_\a\Om\ra=\int_{\D}d\mu(h)\,h([\a]_K)\label{Gen-fun}
\eeq
where $\op{T}_\a:=\op{R}([\a]_K)$.

    The usual type of representation theorems follow from this result.
Specifically, the set of all vectors $\{\op{R}(x)\Om|x\in C\sp\star(\HA)\}$
span (a dense subspace of) the Hilbert space $\H$, and hence this cyclic
representation of the $T$-algebra is unitarily equivalent to one on the
Hilbert space $L\sp{2}(\D,d\mu)$ via the map $\H\map L\sp{2}(\D,d\mu)$
defined by $R(x)\Om\mapsto\twid{x}$. In particular, the cyclic vector $\Om$
is represented by the function $\Om(h):=1$ for all $h\in\D$. The associated
operation of $\op{T}_\a$ on $L\sp{2}(\D,d\mu)$ is then simply
\beq
    (\op{T}_\a\Psi)(h) := h([\a]_K)\Psi(h).     \label{Def-RepT0'}
\eeq
Conversely, any measure $\mu$ on the compact Hausdorff space $\D$ leads to
a representation of the $T$-algebra defined by (\ref{Def-RepT0'}).

    To summarise, {\em every continuous, cyclic representation of\/} $\HA$
{\em is of the following form: The Hilbert space of states is\/}
$L\sp{2}(\D,d\mu)$ {\em for some regular measure\/} $d\mu$ {\em and\/}
$\HA$ {\em acts simply by multiplication as in\/} (\ref{Def-RepT0'}).

    The relation between (\ref{Def-RepT0'}) and the heuristic expression
(\ref{Def-RepT0}) is as follows. To each $SU(2)$ connection $A\in\A$ there
corresponds a multiplicative homomorphism $h_A$ (which depends only on the
gauge-equivalence class of $A$) defined by
\beq
h_A([\a]_K) :=T_\a(A)=\half\Tr({\cal P}\exp\oint_\a A)\label{Def-hA}
\eeq
and extended by linearity and continuity to the whole of
$C\sp\star(\A/\G)$. Thus we have a map $j:\A/\G\map\D$ with
$j([A]_\G):=h_A$. Furthermore, according to the Gel'fand transform
(\ref{Gel-trans}), the (classical) functions $T_\a$ on $\A/\G$ have a
natural extension to $\D$ defined by $T_\a(h):=h([\a]_K)$. If the map
$j:\A/\G\map\D$ were a bijection then (\ref{Def-RepT0'}) would be identical
to (\ref{Def-RepT0}). However, as we will see later, although the map $j$
is one-to-one, it is not surjective and hence $\A/\G$ can be viewed as a
proper subset of the space $\D$ of maximal ideals of $C\sp\star(\HA)$. Thus
(\ref{Def-RepT0'}) is a genuine (and mathematically meaningful) extension
of the heuristic representation (\ref{Def-RepT0}).

    Similarly, we can see how the notion of a ``loop transform''
(\ref{Formal-LT}) arises in our formalism. To each vector $\Psi$ in the
Hilbert space $L\sp{2}(\D,d\mu)$ there corresponds a function $\twid{\Psi}$
of loops defined by
\beq
  \twid{\Psi}(\a) :=\la\op{T}_\a\Om,\Psi\ra =
    \int_{\D}d\mu(h)\,h\big([\a]_K\big)\Psi(h)      \label{LT}
\eeq
where, we have used the fact that $h[\a]_K$ are real-valued.

    The expression (\ref{LT}) can be regarded as a mathematically
well-defined, generalised loop transform from functions of $h$ to functions
of $\a$. Indeed, if the map $j:\A/\G\map\D$ were a bijection then
(\ref{LT}) would be precisely the heuristic transform (\ref{Formal-LT}).
However, as mentioned above, $\A/\G$ is a proper subspace of $\D$, and it
is in this sense that we will refer to (\ref{LT}) as a ``generalised'' loop
transform.

    Finally, note that because $h_\a$ satisfies (\ref{Prop-h1}), the action
(\ref{Def-RepT0'}) of the operators $\op{T}_\a$ can be transformed to the
loop states that are in the image of the transform (\ref{LT}):
\beq
(\op{T}_\b\twid{\Psi})(\a) :=
\int_{\D}d\mu(h)\,h\big([\a]_K\big)\,\big(\op{T}_\b\Psi\big)(h) = \half
\big(\twid{\Psi}(\a\circ\b) +\twid{\Psi}(\a\circ\b\sp{-1})\big)
\eeq
This constitutes a precise version of the action of the holonomy operators
on the ``loop representation'' introduced by Rovelli and Smolin.

    It is tempting to think of the loop transform (\ref{LT}), which enables
one to pass from the ``connection representation'' to the ``loop
representation'', as being analogous to the Fourier transform which enables
one to pass from the position to the momentum representation in elementary
quantum mechanics. While there are indeed some qualitative similarities,
there are also some key differences. To see this, recall first that, in the
case of the Fourier transform, the spaces of functions $f(\vec x)$ and
$\twid{f}(\vec k)$ can be specified a priori---they are both
$L\sp{2}(\R\sp{3})$---and the transform is an isomorphism between them. In
the case of the generalised loop transform, only the space of functions on
$\D$ is specified a priori. We do not have a direct control on the space of
loop functions, independent of the transform. It is clear from (\ref{LT})
that every function $\twid\Psi$ in the image of the transform must satisfy
the algebraic relations: $\sum_{i=1}\sp{n}a_i\twid{\Psi}(\a_i)=0$ if
$\sum_{i=1}\sp{n}a_iT_{\a_i}(A)=0$ for all $A\in\A$. Hence, the transform
cannot be inverted on all loop functions.

    One may think of using these conditions to specify the space of loop
states on which the transform {\em can\/} be inverted. There are, however,
two problems. First, the conditions themselves refer not just to loops but
also to the space $\A$ of connections. Even if one were to ignore this
point, and just use these conditions to single out permissible loop states,
in general the transform may fail to be invertible. As matters stand,
therefore, it seems a little premature to call (\ref{LT}) a ``transform''.
Furthermore, the discussion above makes it clear that the basic type of
representation of the holonomy algebra is on functions on $\D$, rather than
loop functions. Roughly speaking, it is because the holonomy algebra is
constructed from loops that its basic representations arise on functions of
connections.

    In fact, the most natural role for a function of loops is as the {\em
generating functional\/} of a measure on $\D$. More specifically, let
$\Ga:C\sp\star(\HA)\map\R$ be any linear continuous function such that
$\Ga(x\sp\star x)\ge 0$ for all $x\in\C\sp\star(\HA)$. In particular, for
all $a_1\ldots a_n\in\C$,
\beq
 \sum_{i,j=1}\sp{n}\bar{a}_ia_j\big(\Ga(\a_i\c\a_j)+
     \Ga(\a_i\c\a_j\sp{-1})\big)\ \ge 0.        \label{Cond-Ga}
\eeq
Then the Gel'fand-Naimark-Segal construction leads at once to a
representation of $C\sp\star(\HA)$ whose generating functional
(\ref{Gen-fun}) is equal to $\Ga$. We will present a few examples of such
generating functionals in section (\ref{Sec:IID}). From this perspective,
(\ref{Gen-fun}) can be viewed as an analogue of the well-known expression
for a conventional scalar field theory
\beq
 \la\Om,e\sp{i\op{\f}(f)}\Om\ra=\int_{E'}d\mu(\chi)\,e\sp{i\chi(f)}
                                    \label{Gen-fun-QFT}
\eeq
in which the test functions $f$ belong to a topological vector space $E$
and the measure is defined on the topological dual $E'$---typically some
sort of space of distributions. Thus, in making the idea of a loop
transform rigourous, we have also provided one possible answer to the
question of what constitutes the analogue for the non-linear space $\A/\G$
of the topological dual $E'$ of a vector space $E$: it is the space $\D$ of
all maximal ideals in the $C\sp\star$-algebra $C\sp\star(\HA)$.

    Finally, it is worth emphasising that the general features of the
representation theory do not depend on the details of the precise topology
placed on $\HA$. The main effects of this topology are
\be
\item the specific ideal space $\D$ that features in the representation is
determined uniquely by the $C\sp\star$-algebra $C\sp\star(\HA)$, and of
course this depends on the topology placed on $\HA$;

\item whether or not a representation $R:\HA\map B(\H)$ is {\em
continuous\/} clearly depends on the topology on $\HA$.
\ee
\noindent However, in all cases $\D$ is a compact Hausdorff space which
contains $\A/\G$ as a proper subset, and the representation formula
(\ref{Def-RepT0'}), loop transform (\ref{LT}), generating function results,
\etc all still hold.

\subsection{Illustrative examples}\label{Sec:IID}
The representations that we will construct in this sub-section are rather
elementary. While some of them have direct applications to certain
topological field theories, none of them is rich enough to capture the full
dynamics of either Yang-Mills theory or general relativity. The main
motivation behind their construction is to illustrate how the general
representation theory of section \ref{Sec:IIC} works in simple examples.
The main idea in all these examples is to use the Gel'fand-Naimark-Segal
(GNS) construction for the $C\sp\star$-algebra $C\sp\star(\HA)$.

    Let us begin with the simplest case and use it to work up to the more
interesting ones. Fix a connection $A_0$ and consider the functional
$\Ga_{A_0}(\a)$ on $\Lx$ defined by
\beq
  \Ga_{A_0}(\a) := T_\a(A_0),           \label{Def-Ga0}
\eeq
which can then be extended to the vector space $F\Lx$ by demanding that it
be linear. It is clear from (\ref{Def-K}) that the value $\Ga_{A_0}(\a)$ of
$\Ga_{A_0}$ depends only on the $K$-equivalence class $[\a]_K$ of loops to
which $\a$ belongs. Hence $\Ga_{A_0}$ has a well-defined projection to the
factor space $F\L_{x_0}/K\equiv\HA$. Furthermore, since the topology on
$\HA$ is defined using the sup norm on the space of continuous, bounded
functionals on $\A/\G$, it follows that the resulting linear functional on
$\HA$ is continuous. It therefore admits an extension to $C\sp\star(\HA)$,
which we denote again by $\Ga_{A_0}$. Note that this functional is
normalised: $\Ga_{A_0}(1)\equiv\Ga_{A_0}([x_0]_K) = 1$, since $T_{x_0}(A)$
is just the unit function on $\A/\G$. Finally, $\Ga_{A_0}$ is a positive
linear functional on $C\sp\star(\HA)$:
\beqn
\lefteqn{\Ga_{A_0}\Big(\big[\sum_{i=1}\sp{n} a_i\a_i\big]\sp\star\,
\big[\sum_{i=j}\sp{n}a_j\a_j\big]\Big) = \half\sum_{i,j=1}\sp{n} \bar{a}_i
a_j \big(T_{(\a_i\circ\a_j)}(A_0) + T_{(\a_i\circ\a_j\sp{-1})}(A_0)\big)}
                                                    \nonumber\\
& & = \Big(\sum_{i=1}\sp{n}\bar{a}_i\,T_{\a_i}\Big)
      \Big(\sum_{j=1}\sp{n}a_j\,T_{\a_j}\Big) =
\Big|\sum_{i=1}\sp{n}a_i T_{\a_i}\Big|\sp{2}\; \ge 0
\eeqn
where we have used the fact that $T_\a(A_0)$ is real. Hence, $\Ga_{A_0}$
can serve as a generating functional in the GNS construction. The resulting
Hilbert space is given by $L\sp{2}(\D,d\mu)$ where the measure $d\mu$ has
support at just one point, $[A_0]_\G$:
\beq
 \la\Psi_1, \Psi_2\ra := \int_\D d\mu(h)\, \bar{\Psi_1}(h)\,\Psi_2(h)=
        \bar{\Psi_1}([A_0]_\G)\,\Psi_2([A_0]_\G),
\eeq
for all $\Psi_1$ and $\Psi_2$ in $L\sp{2}(\D,d\mu)$. Hence the Hilbert
space is just 1-dimensional. The action of the algebra on states is given
by (cf (\ref{Def-T0'})):
\beq
 \big(\op{T_\a}\Psi\big)(h) = h([\a]_K)\Psi(h)= T_\a(A_0)\Psi(h). \eeq

    It is straightforward to construct finite-dimensional representations
by using convex linear combinations of positive linear functionals of this
type. For our second example therefore let us fix $n$ gauge-inequivalent
connections $A_i$, $i=1\ldots n$, and set
\beq
 \Ga_{\{A_i\}} := \sum_{i=1}\sp{n}c_i\Ga_{A_i},
\eeq
where $c_i$ are real positive constants that sum to 1, and where
$\Ga_{A_i}$ is obtained by replacing $A_0$ by $A_i$ in (\ref{Def-Ga0}). The
measure is now concentrated at just the $n$ points $[A_1]_\G,\ldots
[A_n]_\G$ of $\A/\G$, and the Hilbert space of states is $n$-dimensional.
The loop transform (\ref{LT}) can be performed easily:
\beq
 \twid{\Psi}(\a) := \int_\D d\mu(h)\,h([\a]_K)\Psi(A) =
    \sum_{i=1}\sp{n}c_i\,\Psi([A_i]_\G)\,T_\a(A_i).
\eeq
The loop states $\twid{\Psi}_i(\a)\equiv T_\a(A_i)$ provide a natural
orthonormal basis in the loop representation. These basis vectors are
simultaneous eigenstates of the holonomy operators $\op{T}_\b$ with
eigenvalues $T_\b(A_i)$.

    More interesting examples result from convex superpositions of the
positive linear functionals that arise from gauge-equivalence classes of
connections belonging to finite-dimensional surfaces in $\A/\G$. A natural
class of such surfaces is the moduli space of flat connections; the
resulting representations are then relevant to topological field theories.
For definiteness, for our third example, let us suppose that the underlying
spatial 3-manifold $\Si$ has the topology of a 3-torus $T\sp{3}$, and
analyse the structure of the corresponding moduli space. Recall first that
the gauge-invariant information in a connection can be coded in the
holonomy group it defines (which, in our case is a subgroup of $SU(2)$.) If
the connection is flat, the holonomy around any closed loop depends only on
the homotopy class of the loop, and we have a homomorphism from the
homotopy group of $\Si$ onto the holonomy group of that connection. Now,
since the homotopy group of $T\sp{3}$ is abelian, it follows that the
holonomy group of any flat connection must also be abelian. Hence we
conclude that any flat connection is gauge equivalent to $A
=(\sum_{j=1}\sp{3} a_j\,d\varphi_j)(i\tau\sp{3})$, for some choice of real
constants $a_j$, where $\varphi_j$ are the three angular coordinates on
$\Si$ and $\tau\sp{3}$ = $\left({1\atop 0}{0\atop -1}\right)$ is the third
Pauli matrix. Finally, it is straightforward to check that the connections
that result from $a_j$ and $a_j+n_j$ are gauge equivalent for any choice of
integers $n_j$. Therefore, the moduli space ${\cal M}$ of flat connections
has again the topology of a 3-torus, coordinatised by $a_j\in[0,1]$.

    With this information at hand, we can now define our generating
functional for the GNS construction. Let us set
\beq
 \Ga_{\cal M}(\a) := \int_{\cal M} dV(A)\,\Ga_{A}(\a)
            = \int_{\cal M} dV(A)\,T_\a (A)         \label{Def-GaM},
\eeq
and, for simplicity, choose the obvious volume element on ${\cal M}$;
$dV(A)=\Pi_j da_j$. Again, $\Ga_{\cal M}$, is obtained as a convex
superposition of the continuous, normalised positive linear functionals
$\Ga_{A}$ and is therefore itself a continuous, positive linear functional
on $C\sp\star(\HA)$. Furthermore, since the volume of ${\cal M}$ with
respect to $dV(A)$ is equal to one, $\Ga_{\cal M}$ is normalised. It
therefore defines a representation of $\HA$ via the GNS construction. To
make the resulting representation explicit, let us note that each loop $\a$
on $\Si$ defines three integers, $n_1,n_2,n_3$ that label the homotopy
class to which $\a$ belongs: $n_j$ is the number of times that $\a$ winds
around the $j$-th generator of the homotopy group of $\Si$. Since for any
$[A]_\G$ in ${\cal M}$, the number $T_\a(A)$ depends only on the homotopy
class of $\a$, the generating function $\Ga_{\cal M}$ depends on $\a$ only
through $n_j$.

    Finally, the quantity $T_\a(A)$ can be computed explicitly in terms of
$n_j$ and $a_j$ as $T_\a(A)$ = $\cos\big(2\pi\sum n_j a_j\big)$. Hence, the
expression for the generating functional simplifies to:
\beq
 \Ga_{\cal M}(n_j) = \int_0\sp{1} da_1\int_0\sp{1}da_2 \int_0\sp{1}da_3\,
\cos2\pi(n_1a_1+n_2a_2+n_3a_3).
\eeq
The resulting GNS representation now yields the {\em
infinite-dimensional\/} Hilbert space $L\sp{2}_{+}({\cal M},dV(A))$ of
square-integrable functions on ${\cal M}$ that are even under the
reflection $a_j\mapsto -a_j$. (The odd functions yield another irreducible
representation that arises from a different generating functional.)
Nonetheless, this representation of $\HA$ has a large kernel: if
$\sum_{i=1}\sp{n} c_i T_{\a_i}(A) = 0$ for all $[A]\in{\cal M}$, the
operator $\sum_{i=1}\sp{n} c_i \op{T}_{\a_i}$ is equal to zero.

    Finally, note that the generating functional is invariant under the
action of the group ${\rm Diff}_0(\Si)$ of small diffeomorphisms on $\Si$
(\ie those that are homotopic to the identity map on $\Si$); in fact {\em
each\/} quantum state is left invariant by this action. This is the kind of
generating functional that one hopes to use in non-perturbative quantum
gravity. The loop transform (\ref{LT}) has a particularly attractive form
in this case. We have:
\beq
 \twid{\Psi}(\a) = \int_{\cal M}dV(A)\, T_\a(A)\Psi(A),
\eeq
so that $\twid\Psi(\A)$ depends only on the 3 integers $n_j$, and its
functional form is given simply by:
\beq
 \twid\Psi (n_1,n_2,n_3) = \int_0\sp{1} da_1\int_0\sp{1} da_2 \int_0\sp{1}
da_3\,\cos2\pi(n_1a_1+n_2a_2+n_3a_3)\, \Psi(a_1,a_2,a_3).
\eeq
So now the loop transform is essentially the Fourier transform!

    The main steps of our construction go through even if the topology of
the underlying manifold $\Si$ is more complicated. However, in general, the
structure of the moduli space ${\cal M}$ would be more complicated and one
may not be able to provide explicit expressions for the inner product and
the loop transform. Nevertheless, the general features persist: the
generating functionals analogous to (\ref{Def-GaM})---and all quantum
states---are invariant under ${\rm Diff}_0(\Si)$; up to unitary
equivalence, the representations are insensitive to the choice of the
volume element $dV(A)$ on ${\cal M}$; and, the loop states are functions
only of homotopy classes of loops.

    In spite of the fact that the Hilbert space in the last example is
infinite dimensional, the effective configuration space ${\cal M}$ has only
a finite number of degrees of freedom. For our fourth example, therefore,
we will consider a case in which the number of degrees of freedom is
genuinely infinite. The idea is to construct measures that, roughly
speaking, are concentrated on (the distributional dual to) the space of
abelian connections, and to use the Poincar\'e invariant, Gaussian measure
of the abelian theory. For simplicity, let $\Si$ be the vector space
$\R\sp{3}$ and introduce on it a fixed flat metric of Euclidean signature.
Denote by ${\cal C}$ the space of $SU(2)$ connections $A_a$ which, in some
fixed gauge, are of the type $A_a\sp{(i)}= A_a\de\sp{(i)}_3$, for some
divergence-free 3-form $A_a$ on $\Si$. Clearly, ${\cal C}$ has the
structure of a vector space: it is isomorphic to the space of
gauge-equivalence classes of $U(1)$ connections. We can therefore borrow
ideas from the quantum theory of the free Maxwell field.

    Let us begin by recalling the relevant notions from that theory
\footnote{For details on form factors and the loop transform in the
$U(1)$-case, see \cite{AR91}}.
Given any loop $\a$, a distribution-valued, divergence-free vector density
$F\sp{a}(\a,\x)$ can be defined on $\Si$ by
\beq
 \int d\sp{3}\x\,F\sp{a}(\a,\x)\;\w_a :=\oint_\a dl\sp{a}\,\w_a,
                                        \label{Def-FF}
\eeq
for all test 1-forms $\w_a$ on $\Si$. The quantity $F\sp{a}(\a,\x)$ is
called the $U(1)$-{\em form factor\/} of $\a$ and can be expressed more
directly as $F\sp{a}(\a,\x)= \oint_\a
dl\sp{a}(s)\,\de\sp{(3)}(\x,{\vec\a}(s))$, where $s$ is any parameter along
the loop ${\vec\a}(s)$ in $\Si$. It is useful to introduce a {\em smeared
form factor\/} $F\sp{a}_\e(\a,\x)$ via:
\beq
 F\sp{a}_\e(\a,\x) := \oint_\a dl\sp{a}(s)\,f_\e(\x -{\vec\a}(s))
\eeq
where $f_\e(\vec y) := (2\pi\e\sp{2})\sp{-3/2}(\exp-{|{\vec y}|\sp{2}\over
2\e\sp{2}})$. For each positive value of $\e$, the smeared form factor is a
smooth vector density (which belongs to the Schwartz space) and its limit
as $\e\rightarrow 0$ is the distributional form factor (\ref{Def-FF}). Its
Fourier transform is given by $F\sp{a}_\e(\a,\k) = (\exp
-{\e\sp{2}\k\sp{2}\over 2})\cdot \oint_\a dl\sp{a}(s)\,\exp(-i\k\cdot {\vec
\a}(s))$, which is again an element of the Schwartz space for each positive
$\e$ and each loop $\a$.

    After these preliminaries, we can now define positive linear
functionals $\Ga_{\cal C}$---one for each choice of $\e$---that are adapted
to these abelian connections:
\beq
 \Ga_{{\cal C},\e}(\a):= \exp-{1\over4}\int{d\sp{3}\k \over |\k|}
 |F\sp{a}_\e(\k)|\sp{2}.
\eeq
As before, we can extend this functional by linearity and check, by a
direct calculation, that it is indeed a positive linear functional. In
fact, it is closely related to the vacuum expectation functional in the
Maxwell theory:
\beq
 \Ga_{{\cal C},\e}(\a) = \bra{0}\cos\int d\sp{3}\x\,F\sp{a}_\e(\a,\x)
\op{A}_a(\x) \ket{0},
\eeq
where $\ket{0}$ is the Poincar\'e invariant vacuum, and $\op{A}_a$ is the
operator-valued distribution defined by the divergence-free connection in
the Maxwell theory. (Recall that the trace of the $SU(2)$-holonomy of the
connection $A_a(\x)\de\sp{(i)}_3$ around the loop $\a$ is $2\cos\oint_\a
dl\sp{a}\, A_a$.) The Hilbert space of states provided by the GNS
construction can again be regarded as the space of square-integrable
functionals on $\D$, the measure being concentrated on the distributional
dual of ${\cal C}$. As in the previous example, this representation has a
large kernel: if $\sum_{i=1}\sp{n}c_iT_{\a_i}(A) =0$ for all $A$ in ${\cal
C}$, the operator $\sum_{i=1}\sp{n}c_i\op{T}_{\a_i}(A)$ vanishes. Finally,
the loop transform is well-defined. Furthermore, as in the previous
examples, one can provide a characterisation of the loop states directly,
without any recourse to the transform.

    All these representations are continuous and hence extend to the
$C\sp\star$-completion of $\HA$. There exist, however, some interesting
representations of $\HA$ that fail to be continuous with respect to the
topology we have introduced and which therefore cannot be so extended; as
matters stand, they lie outside the general framework of section
\ref{Sec:IIC}. A particularly interesting one arises from the positive
linear functional
\beq
 \Ga(\a) = \cases{1,& if $[\a]_K = [x_0]_K$ \cr
                  \hfil &\hfil                      \cr
                     0,& otherwise\cr},
\eeq
where, as before, $x_0$ denotes the trivial loop. This generating
functional is diffeomorphism invariant, and hence the resulting Hilbert
space carries a unitary representation of ${\rm Diff}_0(\Si)$. The
analogous representation of the $U(1)$ $C\sp\star$-algebra has the property
that the flux of the electric (or, magnetic) field through any 2-surfaces
is quantised. In the full gravitational case, these representations also
appear to play an important role. At the end of section \ref{Sec:IIB} we
saw that there are reasons to explore alternative topologies on $\HA$. The
existence of these positive linear functionals provides another motivation
for such a step.

\section{STRUCTURE OF THE IDEAL SPACE $\D$}
\label{Sec:Ideal-Space}
The space $\D$ is the domain space of quantum states, and hence, to gain a
further understanding of the quantum theory it is necessary to study its
structure in more detail. We begin in sub-section \ref{Sec:IIIA} by showing
that the space $\A/\G$  of smooth connections modulo gauge transformations
is indeed embedded in $\D$. In \ref{Sec:IIIB}, we show that every element
$h$ of $\D$ defines a representation of the semi-group $\L_{x_0}$ by
$2\times2$ matrices such that $h(\a)$ is given by the trace of the matrix
representing $\a$. If $h$ happens to come from an element of $\A/\G$, the
matrix is just the holonomy about the loop $\a$ defined by a smooth
connection. For a general $h$, however, there is no guarantee that there
exists an underlying smooth connection. In the last two sub-sections, we
explore the structure of $\D-\A/\G$, the space of the ``distributional
analogues of $\A/\G$''. In particular, we construct a large family of these
distributional elements. To what extent this family exhausts $\D-\A/\G$
remains, however, an open question.

\subsection{The role of $\A/\G$}\label{Sec:IIIA}
Let us start by considering the map $j:\A/\G\map\D$, defined above
\footnote{We will sometimes write $h_A$ as $h_{[A]_\G}$ to emphasise the
fact that $h_\A$ depends only on the gauge equivalence class of the
connection $A$.}
as $j([A]_\G):=h_A$, for which we have the following important result.

\noindent
{\bf Theorem}

    The map $j:\A/\G\map\D$ is one-to-one.

\noindent
{\bf Proof}

\noindent
Let $A,B\in\A$ be such that $j([A]_\G)=j([B]_\G)$. This implies that, for
all $\a\in\Lx$,
\beq
         \Tr\big(P_\a(A)\big)=\Tr\big(P_\a(B)\big). \label{PA=PB}
\eeq
For each $A\in\A$, the map $P_A:\Lx\map SU(2)$ defined by
$P_A(\a):=P_\a(A)$ is a homomorphism (\cf (\ref{PaPb=})) from the semigroup
$\Lx$ into the group $SU(2)$. Since loops that are thinly equivalent
possess the same holonomy, this map projects down to give a homomorphism of
the group $\Lx/t$ of thin loops to $SU(2)$. In terms of these
homomorphisms, (\ref{PA=PB}) can be rewritten as
\beq
 \Tr\big(P_A([\a]_t)\big)=\Tr\big(P_B([\a]_t)\big).
                                    \label{TrPA=TrPB}
\eeq
Now suppose that $\Lx/t$ has been given some topological structure so that
it becomes a topological group
\footnote{For example, J. Barrett in \cite{Bar91}.}
and such that, for each $A\in\A$, $P_A$ is a continuous function on
$\Lx/t$. If $\Lx/t$ were compact we could apply to (\ref{TrPA=TrPB}) the
well-known theorem that a representation of a compact topological group is
determined up to unitary equivalence by its trace. However, $\Lx/t$ is an
infinite-dimensional space and is therefore most unlikely to be compact.

\noindent
On the other hand, there is the general result \cite{Kel70} that to any
topological group $G$ there is associated a compact group $\g[G]$ and a
continuous homomorphism $L:G\map\g[G]$ such that any continuous
homomorphism $\f:G\map H$ from $G$ to a compact group $H$ factors through
$\g[G]$, \ie there exists some continuous homomorphism $\th:\g[G]\map H$
such that $\f=\th\c L$. Furthermore, the image of $G$ by $L$ is dense in
$\g[G]$. When applied to the case in hand (remembering that $SU(2)$ is a
compact Lie group) this shows that there exists some compact topological
group $\g[\Lx/t]$, a map $L:\Lx/t\map\g[\Lx/t]$, and a family of continuous
homomorphisms $\th_A:\g[\Lx/t]\map SU(2)$ such that $P_A=\th_A\c L$. Then
(\ref{TrPA=TrPB}) implies that, for all $g$ in the image of
$L:\Lx/t\map\g[\Lx/t]$,
\beq
\Tr\big(\th_A(g)\big)=\Tr\big(\th_B(g)\big).\label{TrthA=TrthB}
\eeq
However, the trace operation is a continuous function on $SU(2)$, and
$\th_A$ is a continuous function on $\g[\Lx/t]$. Hence
$\Tr\c\th_A:\g[\Lx/t]\map\R$ is continuous. Moreover, every complex-valued
continuous function on a compact topological group is uniformly continuous,
and hence, since the image of $\Lx/t$ is dense in $\g[\Lx/t]$, the equality
(\ref{TrthA=TrthB}) can be extended to the whole of the compact group
$\g[\Lx/t]$. But then, by the theorem mentioned above, the equalities of
the traces in (\ref{TrthA=TrthB}) implies that there exists some unitary
matrix $V$ such that, for all $g\in\g[\Lx/t]$,
$\th_B(g)=V\th_A(g)V\sp{-1}$. In particular, this is true for $g=L([\a]_t)$
and hence, since $P_A([\a]_t)=\th_A(L([\a]_t))$, we see that
$P_B([\a]_t)=VP_A([\a]_t)V\sp{-1}$, \ie there exists a unitary matrix V
such that, for all $\a\in\Lx$,
\beq
                 P_\a(B) = VP_\a(A)V\sp{-1}.    \label{P=VPV-1}
\eeq
However, if $\t_\Om(A)$ denotes the transform (\ref{Trans-A}) of $A$ under
the gauge function $\Om$, the holonomy function $P_\a$ transforms as
$P_\a\big(\t_\O(A)\big)$ = $\Om(x_0)P_\a(A)\Om(x_0)\sp{-1}$. Furthermore,
Anandan
\footnote{See also S. Kobayashi, in \cite{Kob54}.}
has shown \cite{Ana83} that if the holonomy maps of a pair of connections
are related as in (\ref{P=VPV-1}) then they are connected by some gauge
transformation. Hence (\ref{P=VPV-1}) implies that $[A]_\G=[B]_\G$, which
proves that $j:\A/\G\map\D$ is one-to-one.    \hfill{\bf QED}

\smallskip
      This theorem shows that $\A/\G$ can be regarded as a subset of $\D$.
This raises two important questions:
\be
\item What is the {\em image\/} of the map $j:\A/\G\map\D$? In particular
    \be\item is it {\em equal\/} to $\D$?
       \item If not, is it a {\em dense\/} subset of $\D$? If so, any
multiplicative homomorphism can be approximated arbitrarily closely by one
of the form $h_{[A]_\G}$, $[A]_\G\in\A/\G$. This is typically the situation
in conventional quantum field theory where the topological vector space $E$
---the classical configuration space---is densely embedded in the dual
space $E'$ of distributions.
    \ee
\item What {\em topology\/} is induced on $\A/\G$ by the Gel'fand spectral
topology on $\D$? Is it anything like a normal function space topology or
is it rather wild, as spectral topologies sometimes tend to be?
\ee
\noindent These questions address directly the issue of the precise form of
the ``distributional'' analogue of members of $\A/\G$, defined to be those
elements of $\D$ that do not belong to the subset $\A/\G$.

       One possible path forward lies in the existence of a different way
of looking at our constructions which throws some light on the form of
$\D$. So far we have not exploited the freedom in the choice of topology
that may be placed on $\A/\G$. However, with respect to any of the standard
topologies the classical $T$-functions are continuous as well as bounded,
so that $\HA\subset C_b(\A/\G)$ where $C_b(\A/\G)$ denotes the set of all
complex-valued, bounded continuous functions on $\A/\G$. Now suppose that
the topology on $\A/\G$ is such that it is a {\em completely regular\/}
space
\footnote{A topological space $X$ is said to be completely regular if, for
each $x\in X$ and each open neighbourhood $U$ of $x$, there is a continuous
function $f:X\map[0,1]\subset\R$ such that $f(x)=0$ and $f$ is identically
one on the complement $X-U$ of $U$ in $X$.}
(this is certainly the case for any of the standard metric topologies on
$\A/\G$). Then a well-known result is that
\beq
             C_b(\A/\G)\simeq C\big(\b(\A/\G)\big)
\eeq
where $\b(\A/\G)$ denotes the Stone-Cech compactification of $\A/\G$. In
particular, this implies that the maximal ideal space of the Banach algebra
$C_b(\A/\G)$ is isomorphic to $\b(\A/\G)$.

    On the other hand,  $C\sp\star(\HA)$ is the $C\sp\star$-algebra
completion of $F\Lx/K\simeq\HA$, and suppose it could be shown that $\HA$
is actually a {\em dense\/} subspace of $C_b(\A/\G)$ (equipped with its
usual sup-norm topology). Then the $C\sp\star$-algebra completion of $\HA$
would be equal to $C_b(\A/\G)\simeq C\big(\b(\A/\G)\big)$, and it would
follow that the space $\D$ of maximal ideals of $C\sp\star(\HA)$ is equal
to $\b(\A/\G)$. This would add considerably to our understanding of the map
$j:\A/\G\map\D$ since it would imply that
\be
\item $\D$ is the Stone-Cech compactification of $\A/\G$;
\item the image of $\A/\G$ in $\D$ is {\em dense\/} in $\D$;
\item the topology induced on the subspace $\A/\G$ is precisely its {\em
given\/} one.
\ee
Hence a key question is whether $\HA$ is a dense subset of $C_b(\A/\G)$.
The assumption that something like this is true is really implicit in the
use of the $T$-variables in the first place, \ie we come now to the
question of the precise meaning of the claim that all gauge-invariant
functions on $\A$ can be written in terms of the $T$-variables.

    First we note that $\HA$ is certainly a {\em separating\/} set of
functions on $\A/\G$. For suppose that $T_\a(A)=T_\a(B)$ for all
$\a\in\Lx$. Then $\Tr\big(P_\a(A)\big)=\Tr\big(P_\a(B)\big)$ and, as shown
in the theorem above, this implies that $A$ and $B$ are gauge-related, \ie
$[A]_\G$ = $[B]_\G$. Hence, if $A$ and $B$ are not gauge-equivalent, there
must exist some $\a$ such that $T_\a(A)\ne T_\a(B)$, which is precisely
what is meant by saying that $\HA$ is a separating set of functions on
$\A/\G$.

    One property of the Stone-Cech compactification is that any bounded
continuous function on the dense set $\A/\G\subset\b(\A/\G)$ has a unique
extension to $\b(\A/\G)$. In particular, this is true for the elements of
the (self-adjoint) holonomy algebra $\HA$. Hence, if the set of all
functions obtained in this way separates the points of $\b(\A/\G)$, the
desired result would follow at once from the Stone-Weierstrass theorem. The
same would apply if $\A/\G$ were a compact space, and hence equal to its
own compactification. However, in any of its typical metric topologies
$\A/\G$ is certainly not compact, or even locally-compact, and in fact all
that can be deduced from the Stone-Weierstrass theorem is that $\HA$ is
dense in $C_b(\A/\G;c)$: the function space equipped with the {\em
compact-open\/} topology, not the sup-norm topology used in the spectral
theory above.

    An extension of the Stone-Weierstrass theorem to completely regular
spaces has been given by Hewitt \cite{Hew47} but this requires the elements
of $\HA$ to separate certain closed subsets (rather than just points) of
$\A/\G$, and it is not obvious how to apply this criterion in our case.
This problem is not trivial since Hewitt showed that in any completely
regular space $X$ which is not compact, there necessarily exists some
family of continuous, bounded functions which separates the points of $X$
and yet which is {\em not\/} dense in $C_b(X)$. In the case of interest we
have no instinctive feeling about this, one way or the other. However, even
if it {\em could\/} be shown that $\HA$ is densely embedded in $C_b(\A/\G)$
with the sup-norm topology, this would not give us much idea of what a
general element of $\D$ actually looks like since the Stone-Cech
compactification of a space is an abstract procedure involving ultrafilters
and hence, ultimately, the axiom of choice. For this reason we will
continue our hunt for distributional analogues of elements of $\A/\G$ in a
more direct way.

\subsection{Matrix representations of elements of $\D$}
\label{Sec:IIIB}
The elements $h_A$ of $\D$ defined by (\ref{Def-hA}) are traces of
$2\times2$ matrices, and we wish now to show that every member of $\D$ can
be written in this way (although there is no guarantee that the matrix is
the holonomy of a smooth connection on $\Si$). This result will be relevant
in the next sub-section where we will search for concrete examples of
``distributional'' multiplicative homomorphisms.

    Every $h\in\D$ generates a real-valued function on $\Lx$, also denoted
$h$, which satisfies the three conditions (\ref{Prop-h1}--\ref{Prop-h3}).
Putting $\b=x_0$ in (\ref{Prop-h1}) shows that
\beq
                     h(x_0)=1
\eeq
while putting $\a=x_0$ gives $h(\b)=\half\big(h(\b)+h(\b\sp{-1})\big)$ so
that, for all $\a\in\Lx$,
\beq
                     h(\a)=h(\a\sp{-1}).      \label{ha=ha-1}
\eeq
Finally, since $h(\a)h(\b)=h(\b)h(\a)$, we have
$h(\a\c\b)+h(\a\c\b\sp{-1})$ = $h(\b\c\a)+h(\b\c\a\sp{-1})$ which, using
(\ref{ha=ha-1}) and the fact that $\b\c\a\sp{-1}=(\a\c\b\sp{-1})\sp{-1}$,
implies that, for all $\a,\b\in\Lx$,
\beq
                     h(\a\c\b)=h(\b\c\a).     \label{hab=hba}
\eeq

       If $h:\Lx\map\R$ satisfies (\ref{Prop-h2}), and if $\a$ and $\b$ are
thinly equivalent, then $h(\a)=h(\b)$, and hence $h$ projects down to give
a function on the group $\Lx/t$ which still satisfies (\ref{hab=hba}). If
this group were {\em compact\/} this would imply that $h(\a)$ is
proportional to the trace of some representation of the group. In fact,
$\Lx/t$ is not compact, but this could be overcome by introducing the
compact group $\g[\Lx/t]$ used in the proof of the earlier theorem.

    A somewhat different approach is to exploit the ideas of Giles
\cite{Gil81} which have the advantage of leading to a specific
representation of the semigroup $\Lx$ whose trace is equal to $h$.

\noindent{\bf Theorem}

    Every element $h$ in the maximal ideal space $\D$ of the
$C\sp\star$-algebra $C\sp\star(\HA)$ can be written as the trace of a
$2\times2$ matrix. More precisely, for each $\a\in\Lx$ there exists a
$2\times2$ complex matrix $V_\a$ such that  $\a\mapsto V_\a$ is a matrix
representation of the semigroup $\Lx$ and $h(\a)=\half\Tr(V_\a)$.

\noindent
{\bf Proof}

\noindent
The first step is to consider once more the free vector space $F\Lx$ but
where the product law is now defined (\cf
(\ref{Def-prod1}--\ref{Def-prod2})) as
\beq
 \Big(\sum_{i=1}\sp{n}a_i\a_i\Big)\Big(\sum_{j=1}\sp{m}b_j\a_j\Big) :=
    \sum_{i=1}\sp{n}\sum_{j=1}\sp{m} a_ib_j\a_i\c\a_j
\eeq
and the adjoint operation (\cf (\ref{Def-star})) is:
\beq
 \Big(\sum_{i=1}\sp{n}a_i\a_i\Big)\sp\star :=
                    \sum_{i=1}\sp{n}\bar{a_i}\a_i\sp{-1}.
\eeq
The identity element $e$ is just the constant loop $x_0$.

    For any given $h\in\D$ (necessarily satisfying
(\ref{Prop-h1}--\ref{Prop-h3})) we define the subspace
\beq
 I_h:=\Big\{\sum_{i=1}\sp{n}a_i\a_i\in F\Lx\big|
\sum_{i=1}\sp{n}a_ih(\a_i\c\b)=0 {\rm\ for\ all\ }\b\in\Lx\Big\}.
\eeq
It is easy to show that $I_h$ is a two-sided ideal in $F\Lx$, and hence we
can construct the algebra $F\Lx/I_h$. The basic relation (\ref{Prop-h1})
shows that
\beq
 \a\sp{2}-2h(\a)\a+e=0 {\rm\ mod\ }I_h     \label{3dum1}
\eeq
which suggests writing this expression in the factorised form
\beq
 \a\sp{2}-2h(\a)\a+e=\big(\a-\l_1(\a)e\big)\big(\a-\l_2(\a)e\big)
                                            \label{3dum2}
\eeq
where the complex quantities $\l_1(\a),\l_2(\a)$ are defined by
\beqn
    \l_1(\a)&:=& h(\a)+ i\big(1-h(\a)\big)\sp\half\\
    \l_2(\a)&:=& h(\a)- i\big(1-h(\a)\big)\sp\half.
\eeqn
Note that both square roots give real numbers because of the boundedness
condition $|h(\a)|\le 1$. For any $\a$ with the property that $\l_1(\a)\ne
\l_2(\a)$, we can construct the following objects
\beqn
\r_1(\a)&:=&\big(\l_1(\a)-\l_2(\a)\big)\sp{-1}\big(\a-\l_2(\a)e\big)\\
\r_2(\a)&:=&\big(\l_2(\a)-\l_1(\a)\big)\sp{-1}\big(\a-\l_1(\a)e\big)
\eeqn
which, by virtue of (\ref{3dum1}--\ref{3dum2}), are projection operators in
the algebra $F\Lx/I_h$:
\beq
 \big(\r_1(\a)\big)\sp{2}=\r_1(\a){\rm\ and\ }
            \big(\r_2(\a)\big)\sp{2}=\r_2(\a)
\eeq
and which are ``orthogonal'' in the sense that $\r_1(\a)\r_2(\a)=
\r_2(\a)\r_1(\a)=0$. Clearly we also have $\r_1(\a)+\r_2(\a)=e$.

    Let us fix once and for all a particular $\g\in\Lx$ for which
$\l_1(\g)\ne\l_2(\g)$. Then a key result of the definitions above is the
equation
\beq
 h\big(\a\r_1(\g)\big)h\big(\b\r_2(\g)\big)= \half
        h\big(\a\r_1(\g)\b\r_2(\g)\big)         \label{3dum3}
\eeq
which holds for all $\a,\b\in\Lx$. Now choose any pair of elements $a,b\in
F\Lx/I_h$ such that $h\big(\r_1(\g)a\r_2(\g)b\big)\ne 0$. Then it follows
from (\ref{3dum3}) that \cite{Gil81} the following $2\times2$ matrix (in
which $\r_1$ and $\r_2$ are both evaluated at the chosen $\g\in\Lx$)
\beq
 V_\a:=\pmatrix{2h(\r_1\a)& h(\r_1\a\r_2b)/h(\r_1a\r_2b)\cr
               2h(\r_1a\r_2\a)&2h(\r_2\a)\cr}
\eeq
satisfies the representation law $V_\a\,V_\b= V_{\a\c\b}$. Furthermore, it
is clear that $h(\a)=\half\Tr(V_\a)$, which proves the desired result
\footnote{If $V_\a$ is to be unitary, the elements $g,g'$ must be chosen
accordingly.}.                                  \hfill{\bf QED}

\smallskip
    Hence we see that for any element $h\in\D$ there exists some $2\times2$
representation $\a\mapsto V_\a$ of $\Lx$ such that $h(\a)=\half(\Tr V_\a)$.
The possibility now arises of constructing distributional analogues of
elements of $\A/\G$ by finding elements $h\in\D$ where the corresponding
matrix $V_\a$ depends on a distributional version of a connection $A$ but
in such a way that the trace still satisfies the defining conditions for
$h$ to be a genuine element of $\D$. We will see in the following
sub-sections how this can be done.

\subsection{Group transformations on $\A/\G$}\label{Sec:IIIC}
We wish now to turn to the problem of finding group transformations on
$\A/\G$. There are two distinct, but related, reasons for this. The first
is that, at least formally, it leads to a set of elements of $\D$ that are
not of the form $h_A$ with $A$ a smooth connection. The main idea is to
extend the definition (\ref{Def-hA}) to ``distributional'' connections of
the form $A+\Th$ where $A\in\A$ is a standard smooth connection and $\Th$
is a distributional, Lie-algebra valued, one-form on $\Si$ that transforms
homogeneously under the operation of the gauge group. This extension is
based on the observation that the right hand side of (\ref{Def-hA}) is
defined on a larger class of one-forms than those contained in $\A$: all
that is necessary is that the integral of $A+\Th$ over $\a$ be
well-defined, and this includes, for example, one-forms $\Th$ that have a
$\de\sp{(1)}$ singularity. The map $A\mapsto A+\Th$ can then be viewed as a
$\G$-equivariant group transformation from connections to (distributional)
connections.

    The second reason for studying transformations of $\A/\G$ is that it
gives a geometrical way of understanding the origin of the
$T\sp{1}$-variables of Rovelli and Smolin. Indeed, the $T$-functions
considered so far depend only on the configuration variables $A$ and must
be supplemented with observables that involve the conjugate momenta $e$
too. In the quantum theory of a system with a finite-dimensional
configuration space $Q$, such observables typically arise from group
transformations on $Q$ with the natural symplectic two-form on $T\sp\star
Q$ being used to associate a function on $T\sp\star Q$ with each vector
field on $Q$ that generates an infinitesimal transformation.

    In the present context this raises the interesting, and mathematically
non-trivial, question of how to find group actions on the space $\A/\G$ of
physical Yang-Mills configurations (or, since this space is
infinite-dimensional, perhaps on its ``distributional dual''). For the
moment we will not specify the internal symmetry group $G$ but let it be an
arbitrary real Lie group---we will see later how, once again, the choice
$G=SU(2)$ is rather special.

    One obvious class of transformations on the space of connections $\A$
is
\beq
         A(x)\mapsto(\l_\f A)(x) := A(x)+\f(x)  \label{A->A+f}
\eeq
where $\f$ is any one-form on $\Si$ taking its values in the Lie algebra
$L(G)$ of $G$. However, the connection $A$ is also subject to the gauge
transformation
\footnote{The expression $\Om(x)d\Om(x)\sp{-1}$ is a symbolic way of
writing $(\Om\sp\star\Xi)(x)$ where $\Xi$ is the $L(G)$-valued
Cartan-Maurer one-form on $G$ and $\Om:\Si\map G$.} (\ref{Trans-A})
\beq
 A(x)\mapsto(\t_\Om A)(x) := \Om(x)A(x)\Om(x)\sp{-1}+\Om(x)d\Om(x)\sp{-1}
                            \label{Trans-A'}
\eeq
and (\ref{A->A+f}) is not equivariant with respect to this operation.
Indeed,
\beq
 \big(\t_\Om(\l_\f A)\big)(x) =
\Om(x)A(x)\Om(x)\sp{-1}+\Om(x)d\Om(x)\sp{-1}+\Om(x)\f(x)\Om(x)\sp{-1}
\eeq
whereas
\beq
 \big(\l_\f(\t_\Om A)\big)(x)=\Om(x)A(x)\Om(x)\sp{-1}
                        +\Om(x)d\Om(x)\sp{-1}+\f(x)
\eeq
so that $\t_\Om\c\l_\f\ne\l_\f\c\t_\Om$ unless $G$ is abelian.

    This lack of equivariance means that the transformations (\ref{A->A+f})
do not map gauge-orbits in $\A$ into gauge-orbits, and hence do not project
down to give maps of $\A/\G$ to itself. In a sense, this is hardly
surprising since the set of all transformations of the form (\ref{A->A+f})
is a topological vector space whereas $\A/\G$ is topologically non-trivial.
Clearly what is needed is some non-abelian group action on $\A$ that
preserves the gauge orbits and hence does yield a map of $\A/\G$ to itself.
We will see now how a large number of such transformations can be
constructed.

    First some notation. Let $L\sp{v}$ denote the left-invariant vector
field on the Lie group $G$ associated with the element $v\in T_eG$, \ie
$(L\sp{v})_g:={l_g}_\star v$ where $l_g:G\map G$ is the usual left
translation operation, $l_g(g'):=gg'$. We recall that the Lie algebra
$L(G)$ of $G$ can be identified with the set of all left-invariant vector
fields on $G$, and that the map $v\mapsto L\sp{v}$ establishes a vector
space isomorphism between $T_eG$ and $L(G)$.

    A key step in our construction is the following theorem.

\noindent
{\bf Theorem}

    Let $\mu:G\map\C$ be any smooth class function
\footnote{A {\em class\/} function on a group $G$ is a function
$\mu:G\map\C$ such that, for all $g,g'\in G$, $\mu(gg'g\sp{-1})=\mu(g')$.
This implies that $\mu(gg')=\mu(g'g)$ for all $g,g'\in G$.}
on the Lie group $G$. Then, for all $v\in T_eG$, $g,g'\in G$, the
left-invariant vector field $L\sp{v}$ satisfies
\beq
L\sp{v}_{gg'g\sp{-1}}(\mu)=L\sp{\Ad_{g\sp{-1}}(v)}_{g'}(\mu)\label{Lsp=}
\eeq
where $L\sp{v}(\mu)$ denotes the function on $G$ obtained by letting the
vector field $L\sp{v}$ act on the smooth function $\mu$, and
$\Ad_g:T_eG\map T_eG$ is defined by $\Ad_g(v):={\ad_g}_\star v$ where
$\ad_g:G\map G$, $\ad_g(g'):=gg'g\sp{-1}$.

\noindent
{\bf Proof}

\noindent
We have
\footnote{We are using the definition of a tangent vector $v\in T_pM$ at a
point $p$ of a manifold $M$ as a local derivation of the ring of smooth
functions on $M$. Thus $v$ is a linear map $v:C\sp\infty(M)\map\R$ such
that, for all $f,g\in C\sp\infty(M)$, we have $v(fg)=f(p)v(g)+g(p)v(f)$.}
$$
     L\sp{v}_{gg'g\sp{-1}}(\mu)=\big({l_{gg'g\sp{-1}}}_\star v\big)(\mu)=
 v\big(\mu\c l_{gg'g\sp{-1}}\big).
$$
But, if $a\in G$,
$$
 \mu\c l_{gg'g\sp{-1}}(a)=\mu(gg'g\sp{-1}a)=\mu(g'g\sp{-1}ag)
            =\mu\c l_{g'}\c\ad_{g\sp{-1}}(a)
$$
so that $\mu\c l_{gg'g\sp{-1}}=\mu\c l_{g'}\c\ad_{g\sp{-1}}$. Therefore
\beqn
 L\sp{v}_{gg'g\sp{-1}}(\mu)&=&v\big(\mu\c l_{g'}\c\ad_{g\sp{-1}}\big)
                =\big({\ad_{g\sp{-1}}}_*v\big)(\mu\c l_{g'})\nonumber \\
&=&\Ad_{g\sp{-1}}(v)(\mu\c
l_{g'})={l_{g'}}_*\big(\Ad_{g\sp{-1}}(v)\big)(\mu)  =
L\sp{\Ad_{g\sp{-1}}(v)}_{g'}(\mu).    \nonumber
\eeqn                                               \hfill{\rm\bf QED}

\smallskip
\noindent To proceed any further it is necessary to concentrate on the
situation in which the structure group $G$ admits some $\Ad\,G$-invariant
inner product on the Lie algebra $L(G)$. We will denote this inner product
by $\la\,,\ra$ and, in particular, write $A\sp{v}:=\la v,A\ra$ for $v\in
T_eG\simeq L(G)$. Note that to be able to recover $A$ from the set of all
$A\sp{v}, v\in T_eG$, it is necessary that $\la\,,\ra$ be non-degenerate.
Note also that, written in terms of the variables $A\sp{v}$, the gauge
transformation (\ref{Trans-A'}) becomes
\beq
 A\sp{v}(x)\mapsto(\t_\Om A\sp{v})(x) :=
         A\sp{\Om(x)\sp{-1}v\Om(x)}(x)+\Om(x)d\Om(x)\sp{-1}.
\eeq

    Now consider the object $L\sp{v}_{P_\a(A)}$ where $P_\a(A)\in G$ is the
parallel transport element defined in (\ref{Def-Pa}), and where $\a$
belongs to $\LS$---the space of {\em all\/} continuous, piecewise-smooth
loops in $\Si$, (not just those that map $0\in S\sp{1}$ to the base point
$x_0$). This group element $P_\a$ has the gauge-transformation property
\beq
         P_\a\big(\t_\Om(A)\big)=\Om(\a(0))P_\a(A)\Om(\a(0))\sp{-1}
\eeq
and hence (\ref{Lsp=}) shows that, for any smooth class function
$\mu:G\map\C$,
\beq
L\sp{v}_{P_\a(\t_\Om(A))}(\mu)=L\sp{v}_{\Om(\a(0))P_\a(A)
\Om(\a(0))\sp{-1}}(\mu)
          =L\sp{\Om\sp{-1}(\a(0))v\Om(\a(0))}_{P_\a(A)}(\mu).
\eeq
Now let $k\in T\sp\star_{\a(0)}\Si$ and consider the transformation
$\l_{(\a,k)}$ defined by
\beq
 \big(\l_{(\a,k)}A\sp{v}_a\big)(x):= A\sp{v}_a(x)+\de\sp{(3)}(x,\a(0))k_a
 L\sp{v}\big(P_\a(A)\big)                 \label{Def-l-trans}
\eeq
which maps $A$ into a distributional connection and where, for
typographical convenience, $L\sp{v}_g(\mu)$ has been written as
$L\sp{v}(g)$. We will return shortly to the distributional nature of this
transformation but for the moment let us concentrate on the key property of
(\ref{Def-l-trans}) which is its equivariance with respect to gauge
transformations:
\beq
\l_{(\a,k)}\big(\t_\Om(A\sp{v})\big)=\t_\Om\big(\l_{(\a,k)}(A\sp{v})\big)
\eeq
for all $\a\in\LS$, $k\in T\sp\star_{\a(0)}\Si$ and $v\in T_eG\simeq L(G)$.
This means that (once suitably smeared) the transformations in
(\ref{Def-l-trans}) project down to give transformations on $\A/\G$.

    The next step is to clarify the group-theoretic content of these
operations. First it should be remarked that the set of all transformations
of the type (\ref{Def-l-trans}) do not themselves form a group since the
composition of two such transformations $\l_{(\a,k)}$, $\l_{(\a',k')}$
cannot  be written in this way. Rather, (\ref{Def-l-trans}) is to be viewed
as an infinitesimal transformation, and hence associated with a
(distributional) vector field $Y_{(\a,k)}$ on $\A$. The crucial question is
whether or not the commutator of any two such fields can be expressed as a
linear combination of fields of this type. If they can then, for example,
the set of all finite linear combinations $\sum_{i=1}\sp{n} a_i
Y_{(\a_i,k_i)}$, $a_i\in\R$, forms a genuine Lie algebra.

    In performing the calculations it is useful to introduce a basis set
$\{E_1,\ldots E_{\dim G}\}$ of $L(G)$ and define $C_{ij}:=\la E_i,E_j\ra$.
Then $A\sp{v}$ = $\la v,A\ra$ = $\la v\sp{i}E_i,A\sp{(j)}E_j\ra$ =
$C_{ij}v\sp{i} A\sp{(j)}$ and hence, if $C_{ij}$ is invertible (which we
will assume) with inverse $C\sp{ij}$, we can write
$A\sp{(i)}=C\sp{ij}A\sp{E_j}$, and the transformation (\ref{Def-l-trans})
becomes
\beq
 \big(\l_{(\a,k)}A\sp{(i)}_a\big)(x):=
A\sp{(i)}_a(x)+\de\sp{(3)}(x,\a(0))k_a C\sp{ij} L\sp{E_j}\big(P_\a(A)\big)
           \label{Def-l-trans'}
\eeq
which corresponds to the vector field
\beq
 Y_{(\a,k)}(A)=\sum_{i,j=1}\sp{\dim G}k_aC\sp{ij}L\sp{E_j}\big(P_\a(A)\big)
                    {\de\over\de A\sp{(i)}_a(\a(0))}
\eeq
expressed as a functional differential operator on functionals on $\A$.

      At this point it is pedagogically useful to restrict our attention to
the case where $G$ is a matrix group and can therefore be written as a
subgroup of $GL(m,\R)$ for some $m$. The obvious example of a class
function on $GL(m,\R)$ (or one of its subgroups) is the trace function
$\Tr:GL(m,\R)\map\R$ defined by $\Tr(g):=\sum_{i=1}\sp{m} g\sp{ii}$. If the
components $g\sp{ij}$ of a matrix $g\in GL(m,\R)$ are regarded as forming a
local coordinate system $x\sp{ij}$ on $G$ (\ie $x\sp{ij}(g):= g\sp{ij}$)
then the well-known expression for the left-invariant vector field
$L\sp{v}_g$ is
\beq
     L\sp{v}_g = \sum_{i,j=1}\sp{m}(gv)_{ij}\left({\partial\over\partial
x\sp{ij}}\right)_g
\eeq
where we have identified the tangent space $T_e\big(GL(m,\R)\big)$ with the
space of all $m\times m$ real matrices. Now, since $\Tr(g)
=\sum_{i=1}\sp{m}x\sp{ii}(g)$, it follows that, for all $g\in GL(m,\R)$,
\beq
 \left({\partial\over\partial x\sp{ij}}\right)_g\Tr = \de_{ij}
\eeq
and hence we have the simple result
\beq
             L\sp{v}_g(\Tr)= \Tr(gv).
\eeq
The basic transformation (\ref{Def-l-trans'}) becomes
\beq
 \big(\l_{(\a,k)}A\sp{(i)}_a\big)(x):=
A\sp{(i)}_a(x)+\de\sp{(3)}(x,\a(0))k_a C\sp{ij} \Tr\big(E_j P_\a(A)\big)
\eeq
while the associated vector field is
\beq
 Y_{(\a,k)}(A)=\sum_{i,j=1}\sp{m} k_aC\sp{ij}\Tr\big(E_j P_\a(A)\big)
                    {\de\over\de A\sp{(i)}_a(\a(0))}.
\eeq

       It is now a straightforward matter to compute the commutator of a
pair of vector fields of this type. In particular, for the case where
$G=SU(n)$ we have $C\sp{ij}(E_i)_{ab}(E_j)_{cd}$ =
$\de_{ad}\de_{bc}-{1\over n}\de_{ab}\de_{cd}$, and a direct calculation
gives
\beqn
  \lefteqn{\[Y_{(\a,k)}, Y_{(\b,l)}\] =
k_a\D\sp{a}(\a(0),\b)\big(Y_{(\a\c\b,l)}-{1\over n}(\Tr P_\a)
                                Y_{(\b,l)}\big)}\nonumber\\
 & &- l_a\D\sp{a}(\b(0),\a)\big(Y_{(\b\c\a,k)}-{1\over n}(\Tr P_\b)
Y_{(\a,k)}\big)                     \label{Com-YY}
\eeqn
where $\D\sp{a}(\a(0),\b)$ is the distribution (in the variable
$\a(0)\in\Si$) defined by \cite{RS90}
\beq
 \D\sp{a}(\a(0),\b):=
\int_0\sp{2\pi}ds\,\de\sp{(3)}(\a(0),\b(s))\dot\b\sp{a}(s).
\eeq
Note that $\b\c\a$ means the composition of the curves $\a$ and $\b$ at
their intersection point $\a(0)$. If they do not intersect at either
$\a(0)$ or $\b(0)$ then the commutator in (\ref{Com-YY}) vanishes. Several
remarks should be made about the result (\ref{Com-YY}).

    1. The algebra in (\ref{Com-YY}) was first obtained by Gambini and
Trias  in their discussion of the canonical quantisation of Yang-Mills
theory \cite{GT86}. Their algebra is that of the commutators of quantum
operators and, as we now see, it follows from the basic transformation law
(\ref{Def-l-trans}) on the classical configuration space.

      2. A priori, the vector field commutator (\ref{Com-YY}) does {\em
not\/} yield a genuine Lie algebra since the right hand side contains the
explicit functions $\Tr(P_\a)$ of the canonical variables $A$. However, we
can always write
\beq
 \Tr(E_iP_\b)=\Tr\big(E_iP_\b(0,u_\a)P_\b(u_\a,2\pi)\big)=
\Tr\big(P_\b(u_\a,2\pi)E_iP_\b(0,u_\a)\big)
\eeq
where $u_\a$ is a value of $u$ such that $\b(u_\a)=\a(0)$, and where
$P_\b(0,u_\a)$ denotes the parallel transport along the loop $\b$ from the
point $\b(0)$ to $\b(u_\a)$. In the special case when the structure group
$G$ is $SU(2)$ the familiar relation $\Tr(A)\,\Tr(B)$ =
$\Tr(AB)+\Tr(AB\sp{-1})$ can be used to expand terms like $(\Tr
P_\a)Y_{(\b,l)}$ to give
\beqn
\lefteqn{\[Y_{(\a,k)},Y_{(\b,l)}\]=\half k_a\D\sp{a}(\a(0),\b)
   \big(Y_{(\a\c\b,l)} -  Y_{(\a\sp{-1}\c\b,l)}\big)}\nonumber\\
& &-\half l_a\D\sp{a}(\b(0),\a)\big(Y_{(\b\c\a,k)}-
Y_{(\b\sp{-1}\c\a,k)}\big)                              \label{Com-YY-SU2}
\eeqn
which {\em is\/} a genuine Lie algebra, albeit still with distributional
structure constants. Thus we see another example of the special role of the
choice $G=SU(2)$: it is the only one of the $SU(n)$ series in which the
vector fields associated with the transformation (\ref{Def-l-trans})
produce a proper Lie algebra.

    3. The algebra (\ref{Com-YY-SU2}) is precisely that obtained by Rovelli
and Smolin for the Poisson-bracket relations of their $T\sp{1}$-variables
(which use the group $SL(2,\C)$). In fact each vector field $Y_{(\a,k)}$
gives rise to a unique $T\sp{1}$ variable according to the usual rule that
associates vector fields with functions on a symplectic manifold.
Similarly, if we let the vector field $Y_{(\a,k)}$ act on the particular
function $T_\b:=\half\Tr(P_\b)$ then, for the special case $G=SU(2)$, we
find
\beq
 Y_{(\a,k)}(T_\b)=k_a\D\sp{a}(\a(0),\b)\big(T_{\a\c\b}-
                                    T_{\a\sp{-1}\c\b}\big)
\eeq
and the right hand side is just the Poisson bracket of the
$T\sp{1}$-variable with the $T$-variable, which re-emphasises the
connection with the work of Rovelli and Smolin. Again, we note the special
role of the group $SU(2)$.

\subsection{The strip variables and distributional elements of $\D$}
\label{Sec:IIID}
The presence of the Dirac $\de$-function on the right hand side of
(\ref{Def-l-trans}) means this basic transformation is not really a map
from $\A$ to $\A$ but rather a map from $\A$ to a space of distributional
connections. To secure a transformation from $\A$ to itself it is necessary
to smear (\ref{Def-l-trans}) in some way. On the other hand, the existence
of the distributional transformation (\ref{Def-l-trans}) suggests the
possibility of constructing ``distributional'' elements of $\D$. We wish to
consider both these topics in the present section.

    The $\de$-function might be removed by ``integrating'' over the loops
$\a$. Note that, at least formally, for any measure $\nu$ on the loop space
$\LS$, and any one-form $\w$ on $\Si$, the object
\footnote{The notation $\Th[A,x)$ is used to emphasise that $\Th$ is to be
thought of as a functional of $A\in\A$ and a function of the point
$x\in\Si$.}
\beq
 \Th\sp{v,\w,\nu}_a[A,x):=\int_{\LS}d\nu(\a)\,\de\sp{(3)}(x,\a(0))
            \w_a(\a(0))L\sp{v}\big(P_\a(A)\big)     \label{Def-Th}
\eeq
behaves under gauge transformations as
\beq
 \Th\sp{v,\w,\nu}_a[\t_\Om A,x)=\Th\sp{\Om(x)\sp{-1}v\Om(x),\w,\nu}_a[A,x)
                    \label{Trans-Th}
\eeq
and hence
\beq
     A\sp{v}_a(x)\mapsto A\sp{v}_a(x)+\Theta\sp{v,\w,\nu}_a[A,x)
                                        \label{Th-Trans-A}
\eeq
is an equivariant map from $\A$ into (a possibly distributional version of)
$\A$.

    This issue of whether, and how, to remove the $\de$-function is related
to the general question of how the Lie algebra should be constructed from
the generators labelled by the pairs $(\a,k)$. The simplest choice is to
take just the set of all finite linear combinations of these generators or,
perhaps, the limit of such in some suitable topology. This corresponds to
the use of a discrete measure on the loop space $\LS$. One might also try a
more general integral over $\LS$, although it is not easy to construct
measures on this space, or to make any a priori decision about what types
of measure are appropriate for our purposes.

    One approach is to seek to smear the loops in such a way that a smooth
function is obtained so that the transformation (\ref{Th-Trans-A}) is an
equivariant map from $\A$ to itself. However, it is also possible that the
space $\A$ should be extended to include genuine distributions. As we have
argued already, this certainly might be expected at the quantum level. This
is particularly relevant if we consider again the space $\D$ and try to use
the transformation (\ref{Th-Trans-A}) to construct elements that are not of
the form $h_A$ where $A\in\A$. That this is indeed possible can be seen
from the following argument.

    If the $\de\sp{(3)}$-function is subjected to a line integral over a
loop we are left with a $\de\sp{(2)}$-singularity. Therefore if the
singular connection on the right hand side of (\ref{Def-l-trans}) is
smeared with a two-dimensional integral over the position of the starting
point $\a(0)$ of the loop $\a$, the resulting (still distributional)
connection $A$ should give rise to a genuine $h_A\in\D$. This is because
the construction of $h_A(\g)$ involves integrating over the loop $\g$,
which removes the remaining singularity. This idea can be given a precise
formulation by considering ``strips'' in $\Si$, \ie maps
$S:S\sp{1}\times(-\e,\e)\map\Si$ (for some $\e>0$) that can be thought of
as one-parameter families of loops. Let $u$ and $v$ denote respectively the
parameter around the circle $S\sp{1}$ and the parameter in $(-\e,\e)$. Then
we define (\cf (\ref{Def-Th}))
\beq
 \Th\sp{S,v}_a[A,x) := \int_0\sp{2\pi}\!du\int_{-\e}\sp\e dv\,
\de\sp{(3)}(x,S(u,v))\dot S\sp{b}_v(u)\dot
S\sp{c}_u(v)\e_{abc}L\sp{v}\big(P_{S_{v,u}}(A)\big)
\eeq
where $S_u:(-\e,\e)\map\Si$ and $S_v:S\sp{1}\map\Si$ are defined by
$S_u(v)=S_v(u):=S(u,v)$, and $S_{v,u}$ is the loop defined by $S_{v,u}(t):=
S_v(u+t\,{\rm mod}\,2\pi)$. Thus we have constructed a rather special
measure $\nu$ on $\LS$ by integrating over a one-parameter family of loops
and over the point on each individual loop at which that loop is deemed to
start.

    Concerning the problem of finding transformation groups of
gauge-equivalence classes of connections we note that (as emphasised in the
context of (\ref{Def-Th}--\ref{Th-Trans-A})) the map $A_a\sp{v}(x)\mapsto
A_a\sp{v}(x)+\Th\sp{S,v}_a[A,x)$ is equivariant with respect to gauge
transformations and corresponds to the (still distributional) vector field
$Y_S$ on $\A$ defined by
\beq
 Y_S:=C\sp{ij}\int_0\sp{2\pi}du\int_{-\e}\sp\e dv\;\dot S\sp{b}_v(u)\dot
S\sp{c}_u(v)\e_{abc}L\sp{E_i}\big(P_{S_{u,v}}(A)\big)
            {\de\over\de A\sp{(j)}_c(S(u,v))}.
\eeq
The vector fields corresponding to a pair of strips $S_1$ and $S_2$ have a
vanishing commutator if $S_1$ and $S_2$ do not intersect. If they intersect
just once the relations are
\beq
 \[Y_{S_1},Y_{S_2}\]=\half\e(S_1,S_2)\left(Y_{S_1\c S_2}
                -Y_{S_1\c S_2\sp{-1}}\right)  \label{Com-YSYS}
\eeq
where $\e(S_1,S_2)=\pm 1$ according to the orientations of the strips, and
where $S_1\c S_2$ denotes the strip formed by all the loops in the parts of
the strips that intersect. If the strips intersect more than once then
(\ref{Com-YSYS}) is replaced by a sum of terms, one corresponding to each
intersection region.

    The Poisson-bracket analogue of this algebra was mentioned by Rovelli
in his recent review \cite{Rov91} and is ascribed by him to Smolin. We
expect this algebra to play a key role in the full quantum theory of any
system with configuration space $\A/\G$. In particular, we note that it is
a {\em genuine\/} Lie algebra with structure constants that are real
numbers, not distributions. Thus, for example, it is meaningful to ask if
the algebra can be exponentiated to give a proper (infinite-dimensional)
Lie group. Note however that any such group would not act on $\A/\G$ since
the vector fields $Y_S$ themselves {\em do\/} still contain distributional
features, even though their commutators do not. The proper role of this
group appears rather to be as a group of automorphisms of the holonomy
algebra $\HA$ and hence, by implication, a group of transformations of the
ideal space $\D$.

    Indeed, from the view point of our study of $\D$ the crucial remark is
that, at least formally, $\Th\sp{S,v}[A,x)$ can be used to form a
non-standard element of this space. Specifically, we define
\beqn
\lefteqn{h_{(A,S)}(\g):=\half\Tr\Big\{{\cal P} \exp\oint_\g
(A+\Th\sp{S}[A])\Big\}}                \nonumber\\
        & &=\half\Tr\bigg\{{\cal P}
\exp\int_0\sp{2\pi}dt\,\dot\g\sp{a}(t)\Big(A_a(\g(t)) \nonumber\\
& &\ \ \ \ +C\sp{ij}E_j\int_0\sp{2\pi}du\int_{-\e}\sp\e dv\; \dot
                 S\sp{b}_v(u)\dot S\sp{c}_u(v)\e_{abc}L\sp{E_i}
                \big(P_{S_{v,u}}(A)\big)\Big)\bigg\}
                    \label{Def-hAS}
\eeqn
which clearly satisfies the boundedness condition (\ref{Prop-h3}). The
algebraic condition (\ref{Prop-h1}) is true by construction, and the
condition (\ref{Prop-h2}) is also true under the assumption that the
topology on $\A$ is such that the classical $T$-variables extend
continuously from functions on $\A$ to functions on the set of
distributional one-forms on $\Si$ of which $\A$ is a dense subset.

    The three conditions (\ref{Prop-h1}--\ref{Prop-h3}) are sufficient to
guarantee that (\ref{Def-hAS}) defines a multiplicative functional on the
holonomy algebra $\HA$ although it is not completely obvious that this
extends to the full $C\sp\star$-algebra completion $C\sp\star(\HA)$ of
$\HA$. However, we believe that this is the case, if for no other reason
than that, with hindsight, is clear that the formal representation
(\ref{Def-RepT0}) has a natural extension to distributional connections of
the type above with
\beq
 (\op T_\a\Psi)([A]_\G,S):=\half\Tr\Big\{{\cal P}
\exp\oint_\a(A+\Th\sp{S}[A])\Big\}\Psi([A]_\G,S)\label{Def-Th-dist}
\eeq
where $A$ is a smooth connection and $S$ is a strip. Indeed, so natural is
the representation (\ref{Def-Th-dist}) that we are inclined to conjecture
that elements of $\D$ of this type are {\em sufficient\/} for the quantum
theory of the $T$-variables, \ie there is some sense in which all
representations of the quantum $T$-algebra can be written in the form
(\ref{Def-Th-dist}).


\section{CONCLUSION}
Let us begin by recalling a few facts from the quantum theory of scalar
fields in Minkowski spacetime. In this case, the algebra of operators is
obtained by smearing the field operators $\hat\f(x)$ with test fields
$f(x)$ which belong to a suitable topological space $E$ of smooth
functions, while quantum states $\Psi(\chi)$ are defined on the {\em
dual\/} $E'$ of $E$; in particular, the argument $\chi$ can be a
distribution. The space $E$ is typically dense in $E'$ in a natural
topology. However, the inner product is defined using measures on $E'$
that are typically concentrated on genuine distributions. The operators
and states---and hence the spaces $E$ and $E'$---``interact'' with one
another to define the generating functional (\ref{Gen-fun-QFT}) which
determines the measure completely.

          The overall framework we have constructed in this paper for
gauge theories and gravity is rather similar, except for a striking
difference that arises in the very first step: the introduction of the
algebra of quantum operators. To maintain manifest gauge invariance, we
chose our basic operators to correspond to traces of holonomies
$\op{T}_\a$ around closed loops $\a$. In effect, this means that we are
using {\em distributions\/}, concentrated on loops, as smearing fields for
operators. This is indeed quite unusual. For example, even in the abelian
case, the Fock representation can not support the $\op{T}_\a$ operators
(without some sort of additional smearing). We have shown that, in spite
of this difference, in theories with real connections, the operators
$\op{T}_\a$ {\em can\/} be given the structure of a genuine
$C\sp\star$-algebra. Furthermore, thanks to the powerful machinery of the
Gel'fand spectral theory, we could obtain the form of its general cyclic
representation. The underlying Hilbert space of states is always
$L\sp{2}(\D,d\mu)$, where $\D$ is the space of maximal ideals of the
$C\sp\star$-algebra and $d\mu$ is a regular measure on this compact,
Hausdorff space. The operators $\op{T}_\a$ act simply by multiplication:
$\big(\op{T}_\a\Psi\big)(h)=h(\a)\Psi(h)$. These results also enabled us to
give a precise mathematical meaning to the ``loop transform'' introduced by
Rovelli and Smolin.

      Since in conventional quantum field theory the domain space of
states is the space of distributions that belong to the dual of the
classical configuration space, and since the configuration space$\A/\G$ of
gauge theories and gravity is a complicated manifold, one of the goals of
this paper was to explore what the distributional dual to $\A/\G$ might be.
We found that there is a well-defined sense in which the ideal space $\D$
plays this role. Thus, the situation with quantum states is rather similar
to that in conventional quantum field theory. Furthermore, the measure that
defines the inner product arises from a generating functional
(\ref{Gen-fun}) whose form is analogous to that of (\ref{Gen-fun-QFT})
encountered in a conventional field theory. Since the generators
$\op{T}_\a$ of our holonomy algebra are labelled by loops, any functional
on the space $\Lx$ of loops satisfying certain conditions (namely
(\ref{Cond-Ga})) defines a generating functional and hence a representation
of our $C\sp\star$-algebra. From the standpoint of the general
representation theory, it appears that this is in fact the most natural
role for functionals of loops in the quantum description.

      We were able to obtain a number of explicit examples of such
functionals. The last three of these are likely to be useful in exploring
certain aspects of topological field theories, Yang-Mills theory and model
systems---such as the one of Husain and \Kuchar \cite{HuK90}---that share
certain features with full general relativity.

    However, these examples are too simple to carry the full content of
the dynamics of Yang-Mills theory or general relativity in 3+1 dimensions.
It is therefore essential to have a better understanding of the general
representation theory and in particular of the structure of the domain
space $\D$ of quantum states. We obtained several results along these
lines. As expected, the classical configuration space $\A/\G$ is embedded
in $\D$, and $\D-\A/\G$ may be regarded as the space of gauge-equivalence
classes of ``genuinely distributional connections''. To gain insight into
these objects, we associated with each 2-dimensional ``strip'' in the
spatial 3-manifold $\Si$, a distributional vector field $\Th\sp{S}$ on
$\A$ and showed that the connections $A+\Th\sp{S}$ lead to well-defined
elements of $\D$. While we have not been able to provide a complete
analysis of the structure of $\D$, nonetheless there are reasons to
believe that the elements of $\D$ that arise from strips may be the only
genuinely distributional connections that are needed in the quantum
theory.

      As noted in section \ref{Sec:Ideal-Space}, however, the holonomy
operators $\op{T}_\a$ correspond only to the configuration variables. It
turns out that the appropriate momentum variables---which must be linear
in $e\sp{a}_i$, the variables conjugate to the connections
$A_a\sp{i}$---are associated precisely with the vector fields $\Th\sp{S}$
on the configuration space $\A/\G$ that are labelled by the 2-dimensional
strips $S$. They are given by:
\beqn
\op{T}_S(A,e) :&=& \int_{\Si}d\sp{3}x\,\Th_a\sp{S,E_i}
                   e\sp{a}_{(i)}              \nonumber\\
                 &=& C\sp{ij}\int_0\sp{2\pi}du \int_{-\e}\sp{\e} dv
                    \dot{S}\sp{b}_v \dot{S}\sp{c}_u \e_{abc}L\sp{E_i}
                    (P_{S_{u,v}}(A)) e\sp{a}_{j}.
\eeqn
These loop-strip variables are complete in an appropriate sense and,
furthermore, are closed under Poisson brackets with the algebra in
(\ref{Com-YSYS}). The Poisson bracket operation involves just gluing and
breaking the loops and strips, and it is quite remarkable that these simple
geometric operations code the entire symplectic structure of Yang-Mills
theory and/or general relativity. The structure of this algebra is so
simple and geometrically pleasing that we expect it to play an important
role in the quantum theory of any system whose configuration space is of
the form $\A/\G$.

      To complete the discussion of quantum kinematics, it is necessary to
find representations of the corresponding loop-strip operator algebra.
Each of them would, by restriction, yield a representation of the holonomy
algebra. Since we now know the structure of the {\it general} representations
of this algebra, it remains to see which of them can also support our momentum
operators. Among the explicit examples discussed in section \ref{Sec:IID},
the first two---the finite dimensional representations---do not do so.
However, it does appear that the remaining three representations {\em
can\/}. If this expectation survives detailed examination one would have,
in particular, a handle on certain non-perturbative aspects of Yang-Mills
theory. We are currently investigating these and other aspects of the
loop-strip algebra. In the next paper in this series, we will report on one
part of this analysis. We will focus on two model systems---2+1-dimensional
gravity and the $U(1)$ (or, Maxwell) theory---and analyse systematically
the structure and the representations of the loop-strip algebras in these
cases.

    In a third paper, we hope to return to 3+1-dimensional general
relativity. In this case, the list of open issues is rather long. First,
there is the problem of reality conditions, \ie, of the $\star$-relations
on $\HA$. One might want to by-pass this temporarily by focussing on
gravitational fields with one (space-like) Killing vector, in which case
the connection is real. However, it now takes values in the Lie-algebra of
$SU(1,1)$---so that the traces of holonomies are unbounded---and is
generically non-flat. Therefore, the strategies that led successfully to a
norm on the holonomy algebra $\HA$ in the $SU(2)$-case as well as in the
case of 2+1-dimensional gravity now fail and one must investigate other
avenues.

    Finally, there is an important and qualitatively different issue that
has remained almost entirely unexplored. In Yang-Mills theory, we have to
worry only about the Gauss constraint, and the strategy of using manifestly
gauge-invariant variables---holonomies $T_\a(A)$ and the momenta
$T_S(A,e)$---is well-suited to handle it. On the other hand, in general
relativity, the corresponding ``kinematical gauge symmetries'' also include
spatial diffeomorphisms, and it is quite unnatural to treat the Gauss
constraint on a different footing from the vector constraints. One
therefore suspects that the loop-strip variables may not be the most
appropriate ones. Is there another set of variables which is invariant not
only under the internal gauge rotations but also under spatial
diffeomorphisms? It would clearly be advantageous to build a quantum
$C\sp\star$-algebra which is based directly on such variables.

\bigskip
\goodbreak
\noindent
{\bf Acknowledgements}

\noindent
This work was motivated by certain key ideas introduced into
non-perturbative canonical gravity by Carlo Rovelli and Lee Smolin. We have
benefited greatly from the numerous discussions we have had with them over
the years. This work was supported in part by the NSF grant PHY 90-16733,
the SERC Senior Visiting Fellowship GR/G3499, and by research funds
provided by Syracuse University.


\end{document}